\documentclass[12pt]{iopart}

\usepackage{graphicx}
\usepackage{amssymb}
\usepackage[mathscr]{eucal}

\newcommand{\email}[1]{\ead{#1}}
\newcommand{\affiliation}[1]{\address{#1}}
\newcommand{\acknowledgments}{\ack}

\newcommand{\el}{\ell_s}

\newcommand{\us}{\mathrm{s}}

\newcommand{\ie}{\textsl{i.e.}~}
\newcommand{\eg}{\textsl{e.g.}~}
\newcommand{\gs}{g_\us}
\newcommand{\ells}{\ell_\us}

\newcommand{\gsimm}{\raise.3ex\hbox{$>$\kern-.75em\lower1ex\hbox{$\sim$}}}
\newcommand{\lsimm}{\raise.3ex\hbox{$<$\kern-.75em\lower1ex\hbox{$\sim$}}}

\def\a{\alpha}
\def\b{\beta}

\def\R{R}
\def\tr{{\rm{tr}}}

\begin{document}

\title{Dirac Born Infeld (DBI) Cosmic Strings}

\author{Eugeny Babichev}
\email{babichev@apc.univ-paris7.fr}
\affiliation{APC (Astroparticules et Cosmologie), UMR 7164 (CNRS,
Universit\'e Paris 7, CEA, Observatoire de Paris), 10, rue Alice Domon
et L\'eonie Duquet, 75205 Paris Cedex 13, France}

\author{Philippe Brax}
\email{philippe.brax@cea.fr}
\affiliation{Institut de Physique
Th\'eorique, CEA, IPhT, CNRS, URA 2306, F-91191Gif/Yvette Cedex, France  and
Institut d'Astrophysique de Paris, UMR 7095-CNRS, Universit\'e Pierre et Marie
Curie, 98bis boulevard Arago, 75014 Paris, France}

\author{Chiara Caprini}
\email{chiara.caprini@cea.fr}
\affiliation{Institut de Physique
Th\'eorique, CEA, IPhT, CNRS, URA 2306, F-91191Gif/Yvette Cedex, France}

\author{J\'er\^ome Martin}
\email{jmartin@iap.fr}
\affiliation{Institut d'Astrophysique de Paris, UMR
7095-CNRS, Universit\'e Pierre et Marie Curie, 98bis boulevard Arago,
75014 Paris, France}

\author{Dani\`ele A.~Steer}
\email{steer@apc.univ-paris7.fr}
\affiliation{APC (Astroparticules et Cosmologie), UMR
7164 (CNRS, Universit\'e Paris 7, CEA, Observatoire de Paris), 10, rue
Alice Domon et L\'eonie Duquet, 75205 Paris Cedex 13, France and
CERN Physics Department, Theory Division, CH-1211 Geneva 23,
Switzerland}

\date{today}

\begin{abstract}
Motivated by brane physics, we consider the non-linear Dirac-Born-Infeld
(DBI) extension of the Abelian-Higgs model and study the corresponding
cosmic string configurations. The model is defined by a potential term,
assumed to be of the mexican hat form, and a DBI action for the kinetic
terms. We show that it is a continuous deformation of the Abelian-Higgs
model, with a single deformation parameter depending on a dimensionless
combination of the scalar coupling constant, the vacuum expectation
value of the scalar field at infinity, and the brane tension. By means
of numerical calculations, we investigate the profiles of the
corresponding DBI-cosmic strings and prove that they have a core which
is narrower than that of Abelian-Higgs strings. We also show that the
corresponding action is smaller than in the standard case suggesting
that their formation could be favoured in brane models. Moreover we show
that the DBI-cosmic string solutions are non-pathological everywhere in
parameter space. Finally, in the limit in which the DBI model reduces to
the Bogomolnyi-Prasad-Sommerfield (BPS) Abelian-Higgs model, we find
that DBI cosmic strings are no longer BPS: rather they have positive
binding energy. We thus argue that, when they meet, two DBI strings will
not bind with the corresponding formation of a junction, and hence that
a network of DBI strings is likely to behave as a network of standard
cosmic strings.
\end{abstract}


\section{Introduction}
\label{sec:introduction}

The (Wilkinson Microwave Anisotropy Probe) WMAP5
results~\cite{2008arXiv0803.0715G,2008arXiv0803.0570H,2008arXiv0803.0732H,
2008arXiv0803.0593N,2008arXiv0803.0586D,2008arXiv0803.0547K} give strong
indication in favour of cosmic inflation over other mechanisms for the
production of primordial fluctuations~\cite{inf25}. Since inflation
generally takes place at high energy, recently there has been a flurry
of activity in constructing models inspired by or derived from string
theory (for recent reviews see \eg
Refs.~\cite{McAllister:2007bg,Burgess:2007pz,Kallosh:2007ig,HenryTye:2006uv}).
In a large category of these models, particularly brane-antibrane
inflation~\cite{Silverstein:2003hf,Alishahiha:2004eh,Chen:2004gc,Chen:2005ad,Lorenz:2007ze,Lorenz:2008je,Lorenz:2008et,Langlois:2008qf,Langlois:2008wt}
and D3/D7 inflation~\cite{Dasgupta:2002ew,Dasgupta:2004dw}, the end
result of the inflationary phase is the creation of D-strings (as well
as potentially F-strings~\cite{Copeland:2003bj}), interpreted from the
four-dimensional point of view as cosmic strings.  Since cosmic strings
are strongly ruled out as the main originator of primordial
fluctuations, the D-(and indeed F-) string tension is severely
constrained and (under certain assumptions) such that $G_{_{\rm
N}}\mu\lesssim 10^{-6}$ in order to preserve the features of the WMAP5
results~\cite{Pogosian:2003mz}. Nevertheless, the existence and possible
detection of the effects of D-strings in the aftermath of an era of
brane inflation could be a testable prediction of string theory.

\par

D-strings themselves have been conjectured to be in correspondence with
the $D$-term strings of supergravity~\cite{Dvali:2003zh}. One of their
remarkable properties is that they satisfy a
Bogomolnyi-Prasad-Sommerfield (BPS) condition, \ie they have no binding
energy and preserve 1/2 of the original supersymmetries. They also carry
fermionic zero modes and are therefore vorton candidates, leading to
possible interesting phenomenological consequences~\cite{Brax:2006zu}.

\par

The identification between $D$-term strings and D-strings has been made
in the low energy limit, when field gradients are small. Inspired by the
case of open string modes which can be effectively described by a
non-linear action of the Dirac-Born-Infeld (DBI) type, in this paper we
construct models of cosmic strings which depart from the low energy
approximation and generalise the Abelian-Higgs model to a non-linear
one. We will call the resulting topological objects `DBI-cosmic
strings', and they are exact solutions of the generalised non-linear DBI
action. The action we consider is very different from others which have
been discussed in the literature,
Refs.~\cite{Moreno:1998vy,Yang:2000uj,Sarangi:2007mj,Brihaye:2001ag,
Babichev:2006cy,Babichev:2007tn}, and in particular does not lead to
pathological configurations. In the limit of small field gradients our
DBI strings reduce to Abelian-Higgs strings.  We construct the DBI
string solutions numerically in a broad range of parameter space, using
two numerical methods: a relaxation method and a shooting algorithm. In
this way, we show, in particular, that DBI strings with a potential term
corresponding to the BPS limit of the Abelian-Higgs model are no-longer
BPS.  More specifically, $\mu_{2n} \geq 2\mu_{n}$, where $\mu _n $ is
the action per unit time and length for a string with a winding number
$n$: the equality only holds in the low-energy limit. Borrowing language
from the standard cosmic string
literature~\cite{Hindmarsh:1994re,Vilenkin:1994,Rubakov}, the strings
are therefore in the type II regime (though the deviations from BPS are
small, in a sense we will quantify). The network of strings produced
will therefore not contain junctions, and all the strings will have the
same tension $\mu_{n=1}$. In the cosmological context we therefore
expect the DBI-string network to evolve in the standard way
\eg~\cite{Bevis:2007gh}, containing infinite strings and loops,
radiating energy through gravitational waves, and eventually reaching a
scaling solution.

\par

The paper is organised as follows. In
section~\ref{subsec:abelianmodel} we recall the
properties of Abelian-Higgs cosmic strings, while their realisation in the
D3/D7 system is discussed in subsection~\ref{subsec:d3d7system}.
In section~\ref{sec:dbics}, we first briefly review the different
non-linear actions which have been put forward so far in the literature.
Then in subsection~\ref{subsec:dbiaction}, we motivate and present our
proposed non-linear action for cosmic strings, which we expect to be
applicable when field gradients are large.
%
In section~\ref{sec:dbisol}, we study the DBI-cosmic strings
solutions analytically and numerically.
In subsection~\ref{subsec:analytical}, we present simple analytical
estimates which allow us to roughly guess the form of the DBI string
profiles and, in subsection~\ref{subsec:numerics}, we compute them
numerically by means of two different methods (shooting and over
relaxation). In section~\ref{sec:conclusions} we briefly summarise
our main findings and discuss our conclusions. Finally, the appendix
gives the full non-linear structure of the DBI cosmic string action.

\section{Abelian-Higgs Cosmic Strings}
\label{sec:abelianhiggscosmicstrings}

\subsection{The Abelian-Higgs model}
\label{subsec:abelianmodel}

We begin by recalling briefly the properties of standard Abelian-Higgs
cosmic strings, and at the same time introduce our notation following
Ref.~\cite{Vilenkin:1994} though we use the signature $(-+++)$.

\par

The Abelian-Higgs model is governed by the action
\begin{equation}
\label{eq:standardaction}
S=-\int {\rm d}^4 x \sqrt{-g}\left[(D_\mu \Sigma)
(D^\mu \Sigma)^{\dagger}+\frac{1}{4}
F_{\mu\nu}F^{\mu\nu}+ V(| \Sigma |) \right] \, ,
\end{equation}
where the potential is given by
\begin{equation}
\label{eq:defpot}
V\left(\left\vert\Sigma \right\vert\right)
= \frac{\lambda}{4}\left(\left\vert \Sigma\right \vert^2
-\eta^2\right)^2\, .
\end{equation}
In Eq.~(\ref{eq:standardaction}), $D_{\mu}$ denotes the covariant
derivative defined by $D_{\mu}\equiv \partial_\mu -iq A_\mu$ with
$A_{\mu }$ the vector potential, $F_{\mu \nu}\equiv \partial _{\mu
}A_{\nu}-\partial _{\nu}A_{\mu}$, and $q$ is the gauge coupling. The
potential is characterised by two free parameters: the coupling
$\lambda>0$ and an energy scale $\eta $.  It is useful to introduce the
dimensionless coupling
\begin{equation}
\beta \equiv \frac{\lambda}{2q^2} = \frac{m^2_{\rm s}}{m^2_{\rm
g}}
\label{beta-def} \, ,
\end{equation}
where the Higgs mass is $m_{\rm s}=\sqrt{\lambda} \eta/\sqrt 2$ and the
vector mass $m_{\rm g}=q \eta$.

\par

\begin{figure}
\begin{center}
\includegraphics[width=7.7cm]{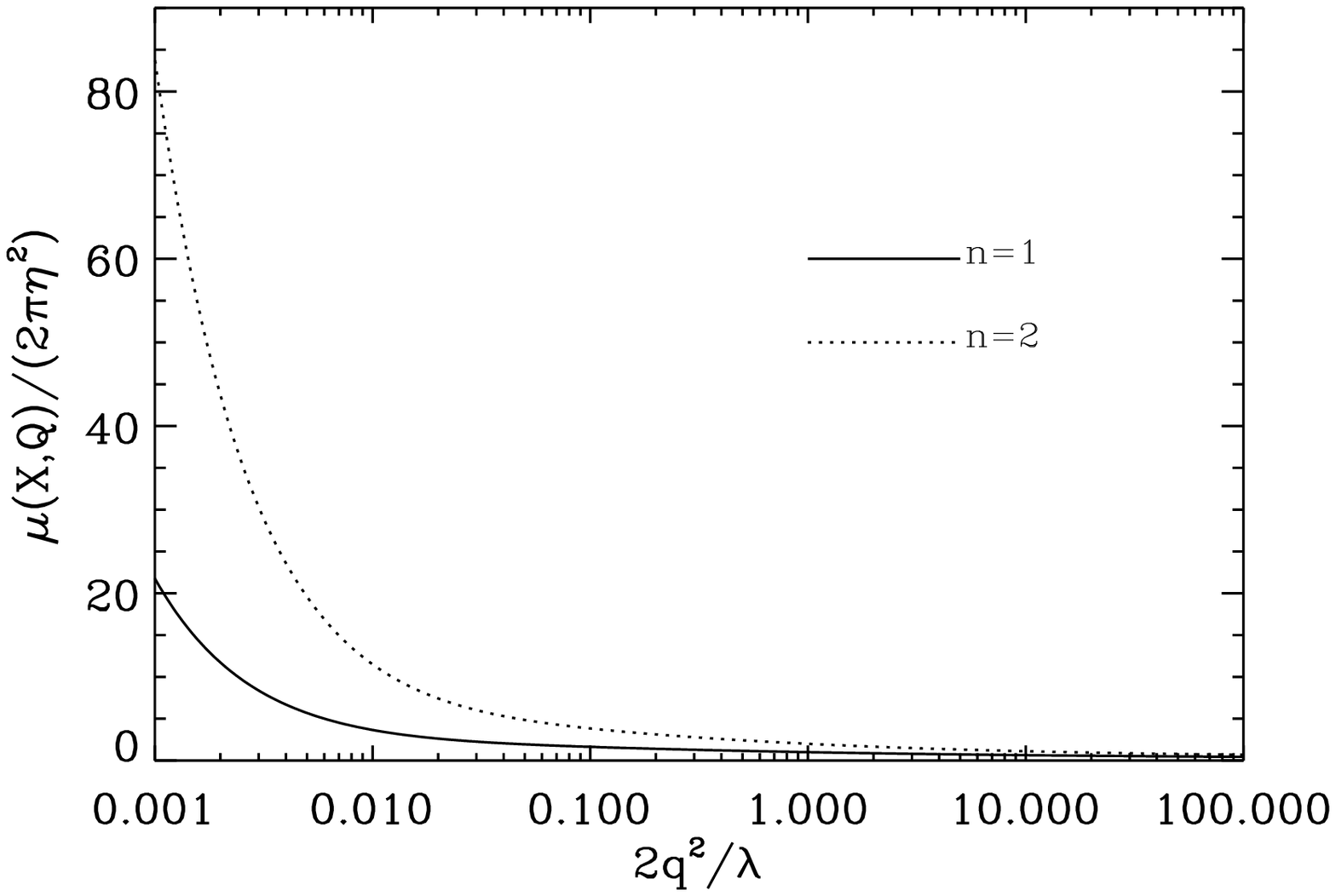}
\includegraphics[width=7.7cm]{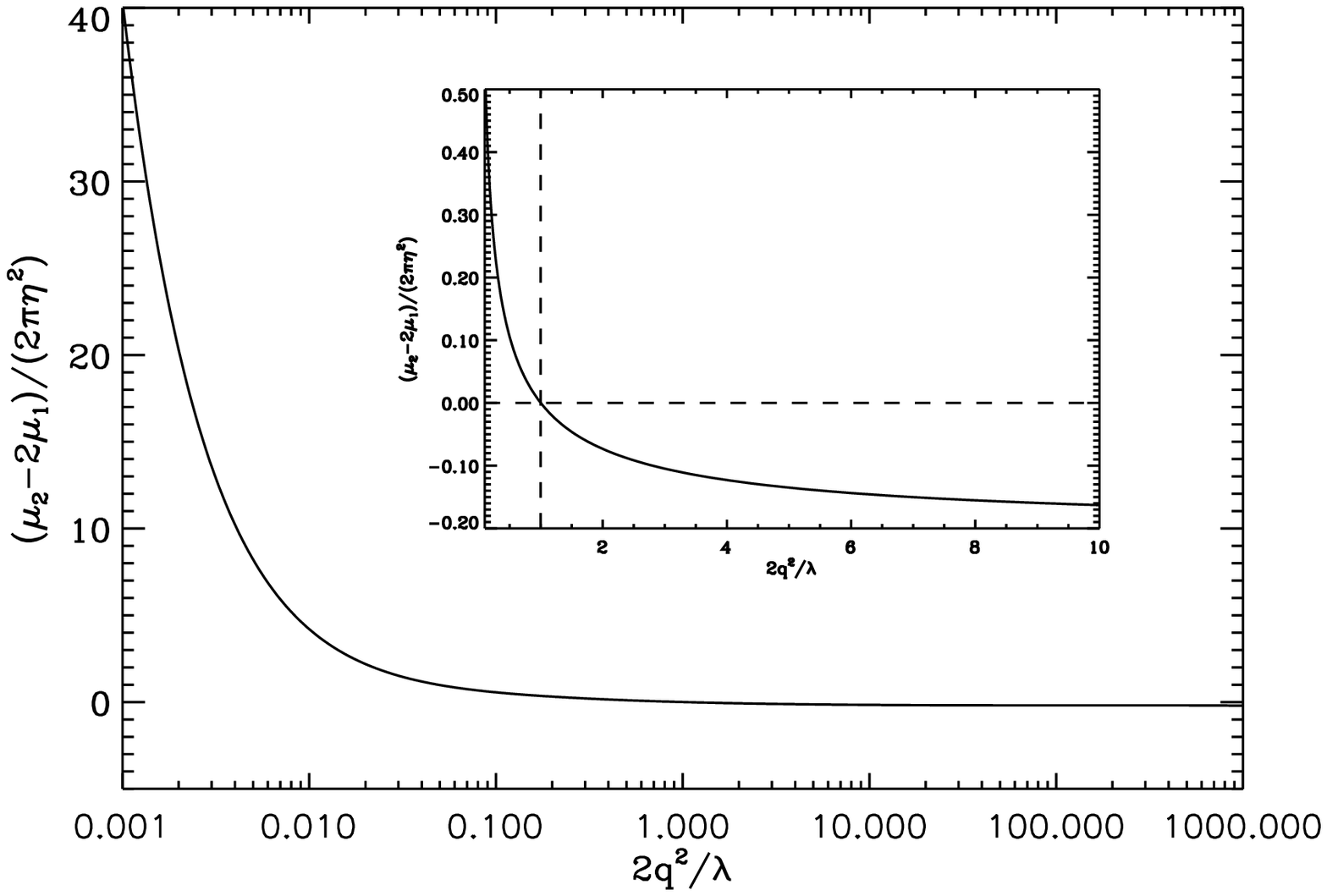} \caption{Left
panel: the tension of $n=1$ and $n=2$ Abelian-Higgs strings as a
function of $\beta^{-1}=2q^2/\lambda $. Right panel: the binding energy
$\mu_{2n}-2\mu_n$ for $n=1$ Abelian-Higgs strings as a function of
$\beta^{-1}$. When the BPS condition $\beta=1$ is satisfied, $\mu
_n=2\pi n\eta ^2$ so that $\mu _{2n}-2\mu _n=0$ as can be verified in
the figure.}  \label{fig:action_vs_charge}
\end{center}
\end{figure}

Due to the non-trivial topology of the vacuum manifold, after gauge
symmetry breaking the model possesses vortex (or cosmic string)
solutions for which the scalar field can be expressed as
\begin{equation}
\label{eq:profileX}
\Sigma \left(r,\theta\right) = \eta
X\left(\rho \right){\rm e}^{in \theta} \, ,
\end{equation}
where we have used the cylindrical coordinates, and the cosmic string is
aligned along the $z$-axis. Here $n$ is the winding number proportional
to the quantised magnetic flux on the string, and we have defined a
rescaled radial coordinate
\begin{equation}
\label{eq:defrho}
\rho \equiv \lambda^{1/2} \eta r \, ,
\end{equation}
with $X\left(\rho \right) \to 1$ at infinity, while $X(0)=0$. In the
radial gauge, the only non-vanishing component of the vector potential
$A_{\mu }$ is $A_{\theta}(\rho)$ with $A_{\theta}(0)=0$. We define
\begin{equation}
Q \equiv n -qA_\theta \, ,
\end{equation}
so that the tension, defined to be the action per unit time and length
$\mu=-S/{\rm d}t {\rm d}z$, can be expressed as
\begin{eqnarray}
\label{eq:actiontx}
\mu_n(\beta)&=&{2\pi {\eta}^2} \int _0^{+\infty}{\rm d}\rho \rho
\left[\left(\frac{{\rm d}X}{{\rm d}\rho}\right)^2
+\frac{Q^2 X^2}{\rho^2}+
\frac{\beta}{\rho^2}\left(\frac{{\rm d}Q}{{\rm d}\rho}\right)^2
+\frac{1}{4}\left(X^2-1\right)^2\right]
\nonumber
\\
&\equiv & 2 \pi \eta^2 g_n\left(\beta^{-1}\right) \, .
\end{eqnarray}
%
The function $g _n\left(\beta^{-1}\right)$ is plotted in
Fig.~\ref{fig:action_vs_charge} (left panel).

In the BPS case, $\beta=1$,
the tension given in Eq.~(\ref{eq:actiontx}) can be re-expressed as
\begin{eqnarray}
\label{eq:reformulaction}
\mu_n &=& {2\pi {\eta}^2} \int _0^{+\infty}{\rm d}\rho  \rho \Biggl\{
\left(\frac{{\rm d}X}{{\rm d}\rho}-\frac{QX}{\rho}\right)^2
+\left[\frac{1}{\rho}\frac{{\rm d}Q}{{\rm d}\rho}
-\frac{1}{2}(X^2-1) \right]^2
\nonumber
\\
& & \qquad \qquad \qquad
+\frac{1}{\rho} \frac{\rm d}{\rm d\rho}\left[Q
(X^2-1)\right]\Biggr\} \, ,
\end{eqnarray}
and is minimised for cosmic strings that are
solutions of the BPS equations
\begin{equation}
\frac{{\rm d}X}{{\rm d}\rho}= \frac{XQ}{\rho}\, ,
\quad
\frac{{\rm d}Q}{{\rm d}\rho}= \frac{\rho}{2} (X^2-1)\, .
\end{equation}
On inserting back into Eq.~(\ref{eq:reformulaction}) one finds
%
\begin{equation}
\mu_n= 2\pi  \eta^2 n \,
\end{equation}
so that $g_n\left(1\right)=n$.
%

The functions $g_{n=1}\left(\beta ^{-1}\right)$ and $g_{n=2}\left(\beta
^{-1}\right)$ are plotted in the left hand panel of
Fig.~\ref{fig:action_vs_charge} whereas the binding energy
$\mu_2-2\mu_1$ is shown in the right hand panel.  In the BPS limit,
$g_n\left(1\right)=n$, and the force between vortices
vanishes~\cite{Jacobs:1978ch}.  For $\beta<1$, the strings are type I
with a negative binding energy, while for $\beta>1$ they are type II
with positive binding energy. Type I string therefore attract and can
form bound states or `zippers'~\cite{Bettencourt:1994kc} linked by
junctions.  Zippers may form (in a certain regime of parameter space)
when two strings in a network collide,
Refs.~\cite{Copeland:2006eh,Salmi:2007ah}.
A network of type II strings, on the other hand,
contains no junctions and the strings all have the same tension
$\mu_{n=1}$.

We now outline how BPS Abelian-Higgs cosmic strings form in certain
string theory models of inflation.  This will motivate our discussion of
DBI strings in section \ref{sec:dbics}.  There we will see that DBI
strings with $\beta=1$ have a positive binding energy and hence repel
each other.

\subsection{BPS Abelian-Higgs strings in the D3-D7 system}
\label{subsec:d3d7system}

In string theory models, cosmic strings can form after the end of
inflation~\cite{McAllister:2007bg,Burgess:2007pz,Kallosh:2007ig,
HenryTye:2006uv}. In the D3/D7
system~\cite{Dasgupta:2002ew,Dasgupta:2004dw} in particular, the two
branes attract during the inflationary period
and then eventually coalesce forming D-strings. The whole picture
(inflation and string formation) can be described in terms of the field
theoretical $D$-term hybrid inflation~\cite{Dvali:2003zh}.  In this
language the D-strings have been conjectured to be analogous to $D$-term
strings, and furthermore --- as we now outline ---
the strings are BPS Abelian-Higgs strings.

\par

In the D3/D7 system there are three complex
fields~\cite{Dasgupta:2002ew,Dasgupta:2004dw}: the inflaton $\phi$ and
the waterfall fields $\phi^\pm$. In string theory, $\phi$ is the
interbrane distance and $\phi^\pm$ are in correspondence with the open
strings between the branes.
In the supersymmetric
language, the K\"ahler potential is
\begin{equation}
K=-\frac{1}{2}\left(\phi-\phi^{\dagger}\right)^2+\vert \phi^+\vert^2
+\vert\phi^-\vert^2\, ,
\end{equation}
leading to canonically normalised fields.  Notice that the
inflation direction is invariant under the real shift symmetry
$\phi\to \phi+c$, thus guaranteeing the flatness of the inflaton
direction at the classical level~\cite{Hsu:2003cy}. The superpotential is
\begin{equation}
W=\lambda \phi\phi^+\phi^- \, .
\end{equation}
During inflation, the $U(1)$ symmetry under which the waterfall fields
have charges $\pm 1$ is not broken, i.e $\phi^\pm=0$. The scalar
potential is flat and picks up a slope at the one loop level. This is
enough to drive inflation. As $\phi$ decreases, it goes through a
threshold after which the waterfall field $\phi^+$ condenses and the
inflaton vanishes. This corresponds to the coalescence of the D3- and
D7-branes. The effective potential describing the condensation is given
by the $D$-term potential (the F-terms all vanish when $\phi=0,\
\phi^-=0$)
\begin{equation}
V_{D} = \frac{g^2}{2}\left(\vert\phi^+\vert^2 -\xi\right)^2
\label{eq:pot}.
\end{equation}
The term $\xi$ is called a Fayet-Iliopoulos (FI)
term~\cite{Binetruy:2004hh}. As $\phi^+$ condenses and $\langle
\phi^+\rangle=\sqrt \xi$, cosmic strings form interpolating between a
vanishing field in the core and $\sqrt \xi$ at infinity. These cosmic
strings are BPS objects preserving one half of the original
supersymmetries. Their tension is known to be $\mu_n= 2\pi n
\xi$~\cite{Dvali:2003zh}. As a consequence,
there is no binding energy as
$\mu_{2n}=2\mu_n$.

\par

In fact~\cite{Dvali:2003zh}, the $D$-term string model of D-string
formation is nothing but an Abelian-Higgs model with particular
couplings
\begin{equation}
{\cal L}= D_\mu \phi^+\left( D^\mu \phi^+\right)^{\dagger}
+\frac{1}{4g^2}F_{\mu\nu}F^{\mu\nu}+V_{D}\, ,
\end{equation}
where $D_\mu \phi^+= (\partial_\mu -i A_\mu )\phi^+$. Upon rescaling
$A_\mu \to A_\mu/g$ and comparing with (\ref{eq:standardaction}), this
leads to the identification
\begin{equation}
q=g\, , \quad \lambda = 2g^2\, ,\quad \xi=\eta^2\, ,
\end{equation}
corresponding to $\beta=1$ and, hence, a BPS system. This explains why
one recovers  $\mu_{2n}=2\mu_n$.

\par

Moreover, the energy scale $\sqrt{\xi}$ can be given a stringy
interpretation. Indeed, one can show that the Fayet-Iliopoulos term is
related to internal fluxes on D7-branes~\cite{Burgess:2003ic}. For this
purpose, let us consider a ten-dimensional metric in the form
\begin{equation}
{\rm d}s^2 =
g_{\mu\nu}{\rm d}X^\mu {\rm d}X^\nu + g_{pq} {\rm d}X^p {\rm d}X^q
+ g_{ij} {\rm d}X^i {\rm d}X^j \, ,
\end{equation}
with $\mu=0,\cdots ,3$, $p=4,\cdots ,7$ and $i = 8,9$. The quantity
$g_{\mu\nu}$ is the four-dimensional metric and $g_{pq}$ and $g_{ij}$
the compactification six-dimensional metric. This corresponds to the
metric on $K^3\times T^2$ compactifications for instance.  The internal
dimensions of the D7-brane are the coordinates $a \equiv (\mu,p)$ while
the D3-brane lies along the $\mu$ coordinates. We denote by $T_7$ the
brane tension and $\gs$ the string coupling. The four-dimensional gauge
coupling is given by
\begin{equation}
\frac{1}{g^2}= \frac{T_7 V_4 \ells ^4}{\gs}\, ,
\end{equation}
where the string length is $\ells= 1/\sqrt {2\pi \alpha'}$ and $V_4=
\int {\rm d}^4 x \sqrt { \det g_{pq}}$ is the volume of the four compact
internal dimensions of the D7-brane. Consider now a dimensionless
magnetic flux $F_{pq}$ along the internal dimensions $(p,q)=4, \cdots,7$,
of a D7-brane.  Then
\begin{equation}
\xi^2=\frac{T_7}{2\gs g^2}\int {\rm d}^4 x \sqrt{-g_{pq}}
F_{pq}F^{pq}\, .
\end{equation}
Notice that the absolute value of the FI term is not fixed, it can be
decomposed as $\xi^2= \zeta/ \el^2$ where the prefactor depends on
$F_{pq}$.

\par

In the following, we will consider cosmic string models for which the
canonical kinetic terms have been replaced by a non-linear term of the
DBI type. In effective actions describing string theory phenomena, and
particularly brane dynamics, such a replacement is mandatory as soon as
the gradient terms in the effective action become large. Indeed, the DBI
action usually describes the dynamics of the open strings in
correspondence with the brane motion (such as the 3-3 and 7-7 open
strings in the D3/D7 system). As we have recalled, the formation of
cosmic strings in the D3/D7 system is governed by the 3-7 strings of no
obvious geometric significance. In such a situation, and assuming that
there could be higher order terms correcting the lowest order
Lagrangian, the effect of the higher order corrections to the kinetic
terms (terms in $\vert D\phi^+\vert^{2p},\ p>1$) would be to induce
modifications of the cosmic string profile and of the tension.

\par

In the following, we do not restrict our attention to a particular
setting such as the D3/D7 system. We discuss possible non-linear
extensions of the Abelian-Higgs system and then motivate a specific,
well defined form.  We then analyse the departure from the BPS case
induced by the higher order terms.

\section{DBI Cosmic Strings}
\label{sec:dbics}

\subsection{Non-standard actions for cosmic strings}
\label{subsec:dbiactioncs}

As discussed in the previous section, we are interested in situations in
which gauged cosmic strings form when higher order corrections to the
kinetic terms in the action cannot be neglected.  In the absence of an
explicit derivation from, say, string theory, we take a phenomenological
approach (which, however, is strongly inspired by string theory).  This
is presented in subsection~\ref{subsec:dbiaction}. We will construct an
action [given in Eq.~(\ref{eq:DBIaction}) and reproduced below] which
satisfies the following two criteria: Firstly, the Abelian-Higgs limit
should be recovered when gradients are small. In particular, the action
should be a continuous deformation of the Abelian-Higgs model. Secondly,
the resulting cosmic string solutions should have no pathological and/or
singular behaviour as the model becomes more and more non-linear (in
field gradients).
In the remainder of this subsection, we compare our action,
Eq.~(\ref{eq:DBIactiondim}):
\begin{eqnarray}
S &\propto &  \int {\rm d}^4x \Biggl\{\sqrt{-\det \left[ g_{\mu
\nu} +({D}_{(\mu}{\phi}) ({D}_{\nu)}
{\phi})^{\dagger}+{F}_{\mu\nu} \right]}-\sqrt{-g} \nonumber \\
& & +\sqrt{-g}  V\left(|{\phi}|\right)\Biggr\} \, .
\end{eqnarray}
to other non-linear actions which have been considered in the literature
and which do not satisfy the above criteria.

\par

In Refs.~\cite{Moreno:1998vy,Yang:2000uj,Brihaye:2001ag} an attempt to
construct a non-linear model for (electrically) charged vortices in
(2+1) dimensions uses an hybrid approach with a (truncated) Born-Infeld
action for the gauge field, a standard linear action for the Higgs
field, and a Chern-Simons term\footnote{No charged vortex solutions
exist when the Chern-Simons term is absent.}:
\begin{eqnarray}
S &=& \int {\rm d}^4x\sqrt{-g}\Biggl[
\sigma^2\Biggl(\sqrt{1+ \frac{1}{2\sigma^2}
F_{\mu\nu}F^{\mu\nu} }-1\Biggr) +\frac{\kappa}{4\pi}
\epsilon^{\mu\nu\rho} A_\mu F_{\nu\rho}
\nonumber \\
& & +\frac{1}{2}\vert D\phi\vert^2 +V(\phi)\Biggr]\, ,
\end{eqnarray}
where $\sigma $ is a parameter of dimension two and $\vert D\phi\vert^2=
(D_\mu \phi) (D^\mu \phi)^{\dagger}$. At a threshold $\sigma=\sigma_{\rm
c}$ corresponding to the very non-linear regime, the gauge field becomes
singular at the origin of the vortex whilst, below the threshold, no
solution exists. Incorporating the Higgs kinetic terms into the square
root, while dropping the Chern-Simons term, leads to the following
expression
\begin{eqnarray}
S &=& \int {\rm d}^4x \sqrt{-g}\Biggl\{
\sigma ^2 \Biggl[\Biggl(1+\frac{1}{2\sigma ^2}F_{\mu \nu}F^{\mu \nu}
+\frac{1}{\sigma^2}\left\vert D\phi\right\vert ^2
+\frac{1}{\sigma ^4}\tilde{F}_{\mu}\tilde{F}_{\nu}D^{\mu}\phi^{\dagger}
D^{\nu}\phi
\nonumber \\ & &
-\frac{1}{2\sigma ^4}\left\vert D \phi\right\vert^2
F_{\mu \nu}F^{\mu
\nu}\Biggr)^{1/2}-1\Biggr]-V\left(\phi\right)\Biggr\}\, ,
\label{tr1}
\end{eqnarray}
where $\tilde{F}_\mu\equiv \epsilon_{\mu\nu\rho} F^{\nu\rho}/2$. With
such a very particular form (which differs from the one we will propose
shortly), one does not find finite energy solutions.

\par

A model for (global) cosmic strings was proposed in
Ref.~\cite{Sarangi:2007mj} with an action given by
\begin{equation}
S=-\int {\rm d}^4x\sqrt{-g}\left[\sqrt{1+ \vert \partial
\phi\vert^2}-1 +V(\phi)\right]\,
\end{equation}
(this is identical to Eq.~(\ref{tr1}) in the global limit). For this
model the solutions become multi-valued and undefined at the origin as
soon as the magnitude of $V\left(\phi\right)$ becomes sufficiently
large~\cite{Sarangi:2007mj}.

\par

In view of these negative approaches where singularities and pathologies
abound, one could be tempted to think that non-linear cosmic string
actions all lead to these problems. Fortunately, a well-behaved action
has been suggested by Sen~\cite{sen} and studied in
Ref.~\cite{Kim:2005tw} in the case of D-strings obtained at the end of
D-${\bar{\rm D}}$ inflation. In such a system, hybrid inflation occurs
and the r\^ole of the waterfall field is played by the open string
tachyon $T$ with a charge $\pm 1$ under the $U(1)$ gauge groups of the
D3- (respectively $\bar{\rm{D}}3$-) brane. When the two branes coincide
the effective action reads
\begin{equation}
S=-T_3\int {\rm d}^4x \, V\left(T,T^{\dagger}\right)\left[
\sqrt {\det{\left( - g^+\right)}}
+\sqrt{\det{\left(- g^-\right)}}\right]\, ,
\end{equation}
where $ g^{\pm}_{\mu\nu}= g_{\mu\nu} \pm \ells^2 F_{\mu\nu}+\left(D_\mu
TD_\nu T^{\dagger} + D_\mu T^{\dagger} D_\nu T\right)/2$ and $V$ is the
tachyon potential.  Since $\det (-
g^+)=\det (-g^-)$ (see the Appendix), the action reduces to
\begin{equation}
S=-2T_3\int {\rm d}^4x V\left(T,T^{\dagger}\right)
\sqrt{\det{( - g^+)}}\, ,
\end{equation}
and it admits BPS vortex solutions.

\par

Topological defects in models with general non-linear kinetic terms were
studied in Refs.~\cite{Babichev:2006cy,Babichev:2007tn}. The action
proposed in~\cite{Babichev:2006cy} for global topological defects
differs from the tachyon action above (in the global limit) as the
potential is added to the generalised kinetic terms:
\begin{equation}
\label{kdefect}
S=\int {\rm d}^4 x\left[M^4 K\left(\frac{X}{M^4}\right) 
- V(\phi)\right]\, .
\end{equation}
Here
\begin{equation}
X\equiv \frac 12(\partial_\mu\phi_a\partial^\mu\phi_a)
\end{equation}
is the standard kinetic term, $K(X)$ is some non-linear function,
$\phi_a$ is a set of scalar fields, $M$ has dimensions of mass, and the
potential term provides the symmetry breaking term. One of the main
restrictions imposed in~\cite{Babichev:2006cy} on the form of the
non-linear function $K(X)$ is that $K(X)$ should have a canonical
asymptotic form, $K(X)\sim X$ as $X\to 0$. However, for large gradients
$K(X)$ could deviate considerably from the canonical kinetic terms. The
former requirement implies a non-pathological behaviour of solutions
far from the defect core, while the different possibilities for $K(X)$
at infinity leads to deviations of the defect from the standard case
inside the core.  The action (\ref{kdefect}) leads to non-pathological
solutions for so-called $k$-defects --- domain walls, cosmic strings and
monopoles --- whose properties can differ considerably from those of
standard defects.

\par

A gauge version of action (\ref{kdefect}) was considered
in Ref.~\cite{Babichev:2007tn}: for a complex scalar field
\begin{equation}
  \label{kvortex}
  S=\int {\rm d}^4 x\left[M^4 K\left(\frac{X}{M^4}\right) 
- V(\phi) - \frac{1}{4}F^{\mu\nu}F_{\mu\nu}\right]
\end{equation}
with
\begin{equation}
X\equiv \frac{1}{2} (D_\mu \phi)(D^\mu \phi)^\dagger.
\end{equation}
It was shown that non-pathological cosmic string solutions exist at
least for some choices of the non-linear function
$K(X)$~\cite{Babichev:2007tn}.

\par

In the following we will motivate and study a non-linear extension of
Abelian-Higgs model which retains some of the properties of the tachyon
and $k$-defect models. The potential will be additive as in the
$k$-defect case while the kinetic terms have a DBI form as in the
tachyon case. However, the kinetic terms are {\it not} a function solely
of $X$ anymore: they differentiate between the radial and angular
gradients of the defects. This springs from the origin of the kinetic
terms as induced from the normal motion of a D3 brane embedded in a
larger space-time.

\subsection{A DBI Action for Cosmic Strings}
\label{subsec:dbiaction}

We now turn to the action that we propose in this article. Consider a
brane model in which cosmic strings appear as deformations of a
brane. (In a sense the brane becomes curved with a puncture at the
location of the string, as we discuss.)  To do so, consider a
ten-dimensional setting as is natural for brane models derived from or
inspired by string theory. We choose a non-warped compactification and
write the ten-dimensional metric in cylindrical form
\begin{equation}
{\rm d}s^2_{10}\equiv g_{AB}^{10} {\rm d}X^A {\rm d}X^B
= {\rm d}s^2_4 + 2g_{\alpha\bar \beta}
{\rm d}Z^\alpha {\rm d}\bar Z^{\bar \beta}\, ,
\end{equation}
where
\begin{equation}
{\rm d}s^2_4\equiv g_{\mu \nu}{\rm d}X^\mu {\rm d}X^\nu
= -({\rm d}X^0)^2+ {\rm d}R^2 + R^2 {\rm d}\Theta^2
+{\rm d}Z^2 \, .
\end{equation}
The metric along the internal dimensions $g_{\alpha \bar \beta}$
($\alpha=5,6,7$) is kept arbitrary, \ie Hermitian and positive
definite, and we have assumed that the six-dimensional manifold is
complex (it could be a Calabi-Yau manifold) therefore having complex
coordinates. The complex coordinates are crucial to analyse cosmic
strings.

\par

Consider the DBI action for a three-brane embedded along the
first four coordinates
\begin{equation}
S=-T\int {\rm d}^4x \sqrt{-\det{\left(\tilde g_{\mu \nu}+\ells^2
{\cal F}_{\mu \nu}\right)}} -\int {\rm d}^4 x\sqrt{-g}
V\left(\sqrt T Z^\alpha\right)\, ,
\end{equation}
where $T$ is the brane tension, ${\cal F}_{\mu \nu}$ is the field
strength on the brane (and has dimension two), distances have dimension
minus one and ${\cal A}_\mu$ has dimension one.  We have included a
potential for the deformations $Z^\alpha$ of the normal directions to
the three-branes.
As suitable when the normal directions are charged under the
world-volume gauge group [in this case the local $U(1)$ on the brane],
we include a covariant derivative in the definition of the induced
metric
\begin{equation}
\tilde g_{\mu \nu}= g_{\mu \nu} + g_{\alpha\bar\beta}\left({\cal D}_\mu
Z^\alpha {\cal D}_\nu \bar Z^{\bar \beta}+{\cal D}_\mu \bar Z^{\bar\beta}
{\cal D}_\nu  Z^{\alpha}\right)\,
\label{hoinduced}
\end{equation}
with
\begin{equation}
{\cal D}_\mu=\partial_\mu -i\hat{q}{\cal A}_\mu \, .
\end{equation}
Clearly, when the gauge fields vanish, $\tilde{g}_{\mu \nu}$ is simply
the induced metric on the brane.
A similar extension of the induced metric to charged fields has already
been introduced in the context of N-coinciding D-branes~\cite{Myers}
with the corresponding non-Abelian $SU(N)$ gauge theory.  There the
brane coordinates are in the adjoint representation and have kinetic
terms involving the $SU(N)$ covariant derivative~\cite{Myers}. We extend
this procedure to the DBI cosmic string situation with a $U(1)$ gauge
group\footnote{In the D-brane context, the brane fields do not carry any
$U(1)$ charge as they belong to the adjoint representation.}

\par

When the six-dimensional metric is nearly flat
$g_{\alpha\bar\beta}=\delta_{\alpha \bar \beta}$ locally, the action
becomes
\begin{eqnarray}
S &=& -T\int {\rm d}^4x \Biggl\{\sqrt{-\det{\left[g_{\mu \nu}
+\left({\cal D}_\mu Z^\alpha {\cal
D}_\nu \bar Z_{\bar \alpha}+{\cal D}_\mu \bar Z^{\bar\alpha} {\cal D}_\nu
 Z_{\alpha}\right)+\ells^2 {\cal F}_{\mu \nu}\right]}}
\nonumber \\ & & \qquad \qquad \qquad
-\sqrt{-g}\Biggr\}
-\int {\rm d}^4 x\sqrt{-g}\, V\left(\sqrt T Z^\alpha\right)\, ,
\end{eqnarray}
where, as usual, we have subtracted the action of the ``flat'' brane so
that the Abelian-Higgs model is recovered when gradients are small.  In
the following we suppose that only one normal direction is excited and
define $\Phi \equiv Z^1$. The resulting action is given by
\begin{eqnarray}
S &=&-T\int{\rm d}^4x \Biggl\{\sqrt{-\det
\left[ g_{\mu \nu} + ({\cal D}_{ ( \mu } \Phi) ({\cal D}_{\nu)}
\Phi)^{\dagger}
+ \el^2 {\cal F}_{\mu\nu} \right] } -\sqrt{-g}
\nonumber \\ & &
\qquad \qquad \qquad
+\sqrt{-g}\frac{V(\sqrt{T}|\Phi|)}{T}  \Biggr\} \, .
\label{eq:DBIaction}
\end{eqnarray}
%
%
Notice that, in the above equation, the potential [\ie $V(x)$ as a
function of $x$] is still given by the expression~(\ref{eq:defpot}) and,
therefore, contains the parameter $\lambda $. When the complex scalar
field $\Phi$ vanishes, Eq.~(\ref{eq:DBIaction}) describes Born-Infeld
electrodynamics~\cite{Born:1934gh}.

\par

We now discuss action~(\ref{eq:DBIaction}) in detail.  In particular we
compare its properties to those of the Abelian-Higgs
action~(\ref{eq:standardaction}) discussed in
section~\ref{sec:abelianhiggscosmicstrings}, and then we construct the
static cosmic string solutions of the action.

\par

A first important property of Eq.~(\ref{eq:DBIaction}) is that, to
leading order in derivatives, it reduces to the standard
action~(\ref{eq:standardaction}) on identifying
\begin{equation}
\Sigma = \sqrt{T} \Phi \, ,
\label{sigmaphilink}
\end{equation}
and redefining the charge according to the following expression
\begin{equation}
q= \frac{ \hat{q}}{\sqrt{T} \ells^2}
\end{equation}
together with the gauge field
\begin{equation}
{\cal A}_\mu=  \frac{A_\mu}{\sqrt{T} \ells^2}\, .
\end{equation}
Hence, if the spatial derivatives characterising DBI-strings are small
(we shall discuss whether or not this is the case below), their
properties should to be very similar to Abelian Higgs strings. More
generally, however, and as discussed in detail in the Appendix where we
calculate the determinant explicitly, Eq.~(\ref{eq:DBIaction}) contains
terms of higher order in covariant derivatives as well as numerous
different mixing terms between ${\cal F}$ and ${\cal D}$ (suitably
contracted). These extra terms could significantly change the string
solution and the resulting strings' properties relative to the Abelian
Higgs case. It follows from this that our action is very different from
that considered by Sarangi in Ref.~\cite{Sarangi:2007mj}, even in the
global case. As a consequence we will find non-pathological cosmic
strings solutions with a continuous limit to Abelian-Higgs strings.

\par

We now focus on the cosmic string solutions of
Eq.~(\ref{eq:DBIaction}). For this purpose, first it is useful to pass
to dimensionless variables, denoted with a hat. Explicitly we set
\begin{eqnarray}
\hat{\Phi}\equiv \frac{\Phi}{\ells}\, ,
\quad
\hat{\cal F}_{\mu \nu}\equiv \ells^2 {\cal F}_{\mu \nu}\, ,
\quad
\hat{\cal D}_\mu\equiv \ells {\cal D}_{\mu}\, ,
\quad
\hat{\eta}\equiv \frac{\eta}{\sqrt{T}\ells}\, ,
\quad
\hat{r}\equiv \frac{r}{\ells}\, ,
\label{dimensionle}
\end{eqnarray}
as well as
\begin{equation}
\hat{V}(|\hat{\Phi}|)\equiv
\frac{\hat{\lambda}}{4} \left(\hat{\Phi}^2 -
\hat{\eta}^2\right)^2\, ,
\end{equation}
where $\hat{\lambda}\equiv\lambda T\ells^4$, so that the action becomes
 \begin{eqnarray}
S &=& -T\ells^4\int {\rm d}^4x \Biggl\{\sqrt{-\det \left[ g_{\mu
\nu} +(\hat{\cal D}_{(\mu} \hat{\Phi}) (\hat{\cal D}_{\nu)}
\hat{\Phi})^{\dagger}+\hat{\cal F}_{\mu\nu} \right]}-\sqrt{-g}
\nonumber \\ & & +\sqrt{-g} \hat V\left(|\hat{\Phi}|\right)\Biggr\}
\, . \label{eq:DBIactiondim}
\end{eqnarray}
We now follow the procedure outlined in section
\ref{subsec:abelianmodel} for Abelian-Higgs cosmic strings, however for
action (\ref{eq:DBIactiondim}). In dimensionless cylindrical
coordinates, ${\rm d}s^2=-{\rm d}\hat t^2+{\rm d} \hat{r}^2
+\hat{r}^2{\rm d}\theta ^2+{\rm d}\hat z^2$ and in the radial gauge
($\hat{A}_{\hat{r}}=0$), the cosmic string profile is
\begin{equation}
\hat{\Phi}=\hat{\eta} X(\rho) {\rm e}^{in\theta}\, ,\quad
Q(\rho)=n-\hat{q}\hat{\cal A}_\theta\left(\hat{r}\right)\, ,
\label{newprof}
\end{equation}
where we have defined a new radial coordinate $\rho $ by the following
expression
\begin{equation}
\rho \equiv{\hat{\lambda}}^{1/2} \hat{\eta }\hat{r}\,
\label{newdef}
\end{equation}
which should be compared to Eq.~(\ref{eq:defrho}).
The boundary conditions on the fields are
\begin{eqnarray}
\lim _{\rho \rightarrow 0} X=0\, , \quad
\lim _{\rho \rightarrow 0} Q=n\, , \quad
\lim _{\rho \rightarrow \infty} X=1\, , \quad
\lim _{\rho \rightarrow \infty} Q=0\, .
\end{eqnarray}
Substituting Eqs.~(\ref{newprof}) and (\ref{newdef}) into
(\ref{eq:DBIactiondim}) as well as using (\ref{dimensionle}), we
find that\footnote{Note that $\gamma^{-2}={\cal D}$ 
where ${\cal D}$ is discussed in the Appendix.}
\begin{eqnarray}
 \left( - \det{g_{\mu \nu}} \right) \gamma^{-2} &= &
-\det \left[ g_{\mu \nu} +(\hat{\cal D}_{(\mu} \hat{\Phi}) (\hat{\cal D}_{\nu)}
\hat{\Phi})^{\dagger}+\hat{\cal F}_{\mu \nu} \right]\, ,
\end{eqnarray}
where we have defined the $\gamma $ factor by
\begin{eqnarray}
\label{eq:defgamma}
\gamma ^{-2}
&\equiv &
\left[1
+\alpha\left(\frac{{\rm d}X}{{\rm d}\rho}\right)^2\right]
\left(1+\frac{\alpha Q^2 X^2}{\rho^2}\right)
+\frac{\alpha}{\rho^2}
\left(\frac{{\rm d}Q}{{\rm d}\rho}\right)^2 \, .
\end{eqnarray}
Hence the string tension defined as $-S/ \ells^2 {\rm d} \hat z{\rm d}
 \hat t$ is given by
\begin{eqnarray}
\mu_n &=& \frac{4\pi {{\eta}^2}}{\alpha}
\int _0^{+\infty}{\rm d}\rho \rho
\left\{\sqrt{
\left[1+ \alpha \left(\frac{{\rm d}X}{{\rm d}\rho}\right)^2\right]
\left(1+\alpha\frac{Q^2 X^2 }{\rho^2}\right)
+\frac{\alpha \beta}{\rho^2}
\left(\frac{{\rm d}Q}{{\rm d}\rho}\right)^2}
\right.
\nonumber \\
& & \left.
\vphantom{\sqrt{
\left[1+ \alpha \left(\frac{{\rm d}X}{{\rm d}\rho}\right)^2\right]
\left[1+\alpha\frac{Q^2 X^2 }{\rho^2}\right]
+\frac{\alpha \beta}{\rho^2}
\left(\frac{{\rm d}Q}{{\rm d}\rho}\right)^2}}
-1+\frac{\alpha}{8}(X^2-1)^2 \right\} \, ,
\label{eq:genten}
\end{eqnarray}
where
\begin{equation}
\label{eq:defalpha}
\alpha \equiv 2\hat{\lambda}\hat{\eta }^4\, ,
\end{equation}
and, as in the Abelian-Higgs case,
\begin{equation}
\beta=\frac{\hat{\lambda}}{2\hat{q}^2}= \frac{\lambda}{2q^2}\, .
\end{equation}
Eq.~(\ref{eq:genten}) is the main result of this section and represents
the non-linear DBI generalisation of the linear Abelian-Higgs model: it
should be compared to Eq.~(\ref{eq:actiontx}). Notice that it involves
the single additional parameter $\alpha$ which measures the deformation
from the Abelian-Higgs model, since Eq.~(\ref{eq:genten}) reduces to the
tension of Abelian Higgs strings in the linear limit $\alpha \to 0$.  As
discussed in section \ref{subsec:abelianmodel}, Abelian-Higgs strings
are BPS when $\beta=1$, and hence for $\beta=1$, DBI-cosmic strings with
tension given by Eq.~(\ref{eq:genten}) are a continuous deformation of
the BPS Abelian-Higgs strings. This property is, of course, very
important and constitutes an additional motivation for the action given
in Eq.~(\ref{eq:DBIactiondim}).

\par

Finally, we note that the argument of the cosmic string profile $\rho$
is also identical to its counterpart in the Abelian-Higgs case, whatever
the value of $\alpha$. Hence we will be able to find continuous
deformations of the cosmic string profiles parameterised by $\alpha$ and
depending on the universal variable $\rho$.

\section{DBI String Solutions}
\label{sec:dbisol}

\subsection{Analytical Estimates}
\label{subsec:analytical}

Having established the model and its action, we now turn to the
solutions of the equations of motion. The DBI cosmic string equations
follow from Eq.~(\ref{eq:genten}) and read
\begin{eqnarray}
& & \frac{{\rm d}}{{\rm d}\rho}\left[\rho \gamma
\left(1+\frac{\alpha Q^2 X^2}{\rho^2}
\right)\frac{{\rm d}X}{{\rm d}\rho}\right] =
\frac{\rho}{2}(X^2-1)X
+\gamma \frac{Q^2X}{\rho}\left[1+\alpha
\left(\frac{{\rm d}X}{{\rm d}\rho}\right)^2
\right]\, ,\nonumber
\label{eq:scalar}
\\ \\
& &\frac{{\rm d}}{{\rm d}\rho}\left(\frac{\gamma}{\rho}
\frac{{\rm d}Q}{{\rm d}\rho}\right)
=\frac{\gamma Q}{\beta \rho }
\left[1+\alpha \left(\frac{{\rm d}X}{{\rm d}\rho}\right)^2
\right]X^2 \, ,
\label{eq:gauge}
\end{eqnarray}
for the scalar field and gauge fields, respectively, where $\gamma $ is
defined in Eq.~(\ref{eq:defgamma}). In the Abelian Higgs limit,
$\alpha=0$, Eqs.~(\ref{eq:scalar}) and (\ref{eq:gauge}) reduce to the
standard cosmic string field equations for which $\gamma=1$. Deviations
from Abelian Higgs strings will occur if $\gamma <1$. Notice that
here the fields are purely space-dependent.  For time-dependent fields,
and particularly in DBI inflationary cosmology with inflaton $\phi(t)$
whose dynamics is described by action (\ref{eq:DBIactiondim}) in the
global limit, then $\gamma$ is a generalisation of the cosmological
Lorentz factor. Indeed, as can be seen from Eq.~(\ref{eq:defgamma}) in
the case when $g_{\mu \nu}$ describes an homogeneous and isotropic
manifold, $\gamma^{-2} =1-{\dot{\phi}^2}/{T(\phi)}$
where $T(\phi)$ is related to the metric of the extra-dimensions. The
difference in sign between spatial and temporal derivatives is
responsible for the fact that deviations from standard cosmology
($\gamma =1$) occur here when $\gamma \rightarrow +\infty$ rather than
$\gamma \ll 1$.

\par

Unfortunately, as is clear from Eqs.~(\ref{eq:scalar})
and~(\ref{eq:gauge}), the DBI cosmic string equations cannot be solved
exactly. We have therefore carried out a full numerical integration of
the equations of motion. For convenience, we will focus on the deformed
BPS case where $\beta=1$ and $\alpha\ne 0$ (though $\beta\neq 1$ and
$\alpha\ne 0$ can also been done with the numerical methods used here).

\par

Before discussing the numerical results, we analyse the asymptotic
behaviour of the fields in order to obtain a rough understanding of the
solution. We will consider the two limits $\rho\to 0$ and
$\rho\to\infty$ and will address two issues. The first one is the
non-existence of singularities in the core of the cosmic string. The
second one will be the determination of the shape of the cosmic string
profile both at the origin and at infinity. In particular we will find
that the functional form of the string profile is similar to the
Abelian-Higgs case inside the core, the only difference springing from
$\alpha$-dependent factors.

\par

Consider first the $\rho \rightarrow 0$ limit and let us examine the
possibility of singular DBI strings deep in the string core.  As already
discussed, the DBI features of the solutions depend on $\gamma$. In
particular, extreme deviations from the Abelian-Higgs case would appear
if $\gamma \to 0$ at the origin. This can only happen if the derivative
of $X$ and/or $Q$ become extremely large, \ie the string becomes
singular. Let us first assume that the gradient of $Q$ becomes large and
dominates the $\gamma$ factor, \ie $\gamma\sim \rho {\rm d}\rho/
\left(\sqrt{\alpha}{\rm d}Q\right)$. The gauge equation becomes
non-sensical as the left-hand of (\ref{eq:gauge}) vanishes and the
right-hand side does not. Hence there is no regime where the gradient
$Q$ is arbitrarily large leading to $\gamma\to 0$ at the origin. We now
examine the possibility that $X$ becomes singular at the origin with a
large gradient. In this limit, we find
\begin{equation}
\gamma \sim \frac{\rho}{\alpha X\left({\rm d}X/{\rm d}\rho\right)}
\left[1-
\frac{1}{2}\frac{1}{\alpha \left({\rm d}X/{\rm d}\rho\right)^2}
-\frac{1}{2n^2}\frac{\rho ^2}{\alpha X^2}\right]\, ,
\end{equation}
where $Q\sim n$ close to the origin and we have expanded $\gamma$ in
$1/\left[\alpha \left({\rm d}X/{\rm d}\rho\right)^2\right] \ll 1$ and
$\rho^2/\left(\alpha X^2\right) \ll 1$, this last condition being the
only one compatible with the condition on the derivative of $X$. Working
to first order in these two parameters, the profile is determined by
\begin{equation}
\frac{{\rm d}}{{\rm d}\rho}\left\{X
\left[\frac{1}{n^2}\frac{\rho ^2}{\alpha X^2}
-\frac{1}{\alpha \left({\rm d}X/{\rm d}\rho\right)^2}\right]\right\}
=\frac{{\rm d}X}{{\rm d}\rho}\left[
\frac{1}{\alpha \left({\rm d}X/{\rm d}\rho\right)^2}
-\frac{1}{n^2}\frac{\rho ^2}{\alpha X^2}\right]\, .
\end{equation}
Notice that to zeroth order in the two small parameters, the equation is
tautological.  In the limit $\rho \to 0$ with an ansatz $X\sim
\rho^\delta$ the equation of motion is satisfied for $\delta^2=n^2$. The
only solution satisfying $X(0)=0$ is obtained for $\delta=n$ which has
finite derivative at the origin. This contradicts our premises and, as a
result, we conclude that singular DBI strings do not exist.

\par

Having shown that the strings are not singular, we will now show that
the functional form of the solutions is similar to the ones in the
Abelian-Higgs case. Let us assume that, in the limit $\rho\to 0$, the
DBI solutions are of the form
\begin{equation}
X(\rho)=A_{_{\rm DBI}}\rho^p \, , \quad Q(\rho)=n-B_{_{\rm
DBI}}\rho^q\, , \label{asympt0}
\end{equation}
where $p$ and $q$ are two constants which we will determine below, while
$A_{_{\rm DBI}}$ and $B_{_{\rm DBI}}$ are two constants to be obtained
by numerical integration; and we have taken into account the boundary
conditions at $\rho=0$: $X(0)=0$ and $Q(0)=n$. By direct substitution of
the asymptotic form (\ref{asympt0}) into the equations of motion
Eqs.~(\ref{eq:scalar}), (\ref{eq:gauge}) and taking the limit $\rho\to
0$, one can check that the correct asymptotic form for $X$ and $Q$
reads,
\begin{equation}
X(\rho)=A_{_{\rm DBI}}\rho^n \, , \quad Q(\rho)=n-B_{_{\rm
DBI}}\rho ^2\, . \label{XQ0}
\end{equation}
Thus $p=n$ and $q=2$ and, as guessed above, the only difference between
DBI cosmic strings and Abelian-Higgs cosmic strings close to the origin
is in the numerical values of the prefactors $A_{_{\rm DBI}}$ and
$B_{_{\rm DBI}}$ which are $\alpha$-dependent. In particular, these
coefficients become large for large $\alpha$ implying that away from the
origin but for reasonable and finite values the gamma factor becomes
noticeably different from one, \ie the cosmic strings are in a mild DBI
regime.

\par

From Eqs.~(\ref{eq:defgamma}) and~(\ref{XQ0}) it immediately
follows that $\gamma$ is always finite at $\rho=0$. This a salient
point as it confirms that the cosmic strings constructed with a
DBI action are non-singular at the origin. This is of course yet
another argument supporting the fact that Eq.~(\ref{eq:genten})
represents the natural DBI generalisation for cosmic strings.

\par

More precisely we find that as $\rho\to 0$ there are two possible
regimes: the standard regime where $|1-\gamma|\ll 0$ and a mild
DBI regime where $\gamma < 1$ but finite. Let us first analyse the
global string, for which $Q=0$. It is clear from
Eqs.~(\ref{eq:defgamma}) and~(\ref{XQ0}) that for $n\geq 2$ the
mild DBI regime cannot be realised around $\rho\to 0$. Note,
however, that for $\alpha\gg 1$ we find numerically that $A_{_{\rm
DBI}}\gg 1$, which implies that away from the origin the gradient
${\rm d}X/{\rm d}\rho$ becomes large, so that the solution is in
the mild DBI regime. For $n=1$ the mild DBI regime is valid
starting from $\rho=0$, if $\alpha\gg 1$. For $\alpha\ll 1$ the
regime is always of non-DBI type, independently of $n$.

\par

In the case of gauge strings, the situation is similar in the
limit $\rho\to 0$. Again for $\alpha\ll 1$ the non-DBI regime is
realised. For $\alpha\gg 1$ we find numerically that $A_{_{\rm
DBI}}$ and $B_{_{\rm DBI}}$ in Eq.~(\ref{XQ0}) are large. Thus the
gradient ${\rm d}Q/{\rm d}\rho$ is large too, while the terms
proportional to ${\rm d}X/{\rm d}\rho$ and to $Q^2X^2$ are large
only for $n=1$, and they are small in a small region around
$\rho=0$ for $n\geq 2$. However, these terms become large away
from the origin, since a large constant $A_{_{\rm DBI}}$ implies
that ${\rm d}X/{\rm d}\rho$ becomes large at some point. In
conclusion, we find that for $\alpha$ large enough, the cosmic
strings are in a mild DBI regime for finite values of $\rho$. This
is confirmed numerically.

\par

Finally let us notice that at infinity, independently of $\alpha$, both
the gradients ${\rm d}X/{\rm d}\rho$ and ${\rm d}Q/{\rm d}\rho$ are
small, and the cosmic string matches the standard behaviour. This is in
agreement with the general findings for topological defects with a
non-canonical kinetic term.

\par

In summary, the difference between the Abelian-Higgs and DBI
strings will be small very far from the core of the string, while
the DBI string can differ from the Abelian-Higgs one inside the
core of the string:  the larger $\alpha$ the larger  the
difference inside the core.

\subsection{Numerical Solutions}
\label{subsec:numerics}

As mentioned above, the equations of motion~(\ref{eq:scalar})
and~(\ref{eq:gauge}) cannot be solved analytically. For this reason, we
now turn to a full numerical integration.

\par

As is well-known, the numerical integration is non-trivial because the
boundary conditions are not fixed at the same point. The solutions
discussed is this article have been obtained by means of two independent
methods: a relaxation method~\cite{AP84,Peter:1992dw,Ringeval:2002qi}
and a shooting method. More precisely, the former is in fact the over
relaxation method. The over relaxation method differs from the
relaxation method (also known as the Newton iteration method) by the
fact that the Newtonian iteration step is multiplied by a factor of
$\omega$. In the standard case, convergence for the over relaxation
method is guaranteed provided the over relaxation parameter $\omega <2$
and, therefore, a good choice is for instance $\omega \sim 1.99$. Here,
the highly non-linear nature of the equation of motions may render the
over relaxation method unstable. To deal with this problem, we have
considered a ``step-dependent'' over relaxation parameter $\omega $
interpolating from $\omega \sim 1$, close to the origin, and to $\omega
\to 1.3 $ at ``infinity''. As already mentioned the choice $\omega
=1.3\ll 1.99$ is due to the highly non linear behaviour of the
equations. We have observed very severe instabilities for higher values
of $\omega$. On the other hand, the shooting method can be directly
implemented in its standard formulation in the case of global strings,
since there is only one integration constant to be obtained, $A_{_{\rm
DBI}}$. While in the gauge case the presence of two ``shooting''
constants, $A_{_{\rm DBI}}$ and $B_{_{\rm DBI}}$, makes the direct
implementation of the standard scheme impossible, we have thus modified
the shooting method appropriately. All in all, the two different
numerical procedures, a relaxation and a shooting method, give the same
numerical solutions, up to small numerical errors.

\par

Numerical integration of the equations of motion~(\ref{eq:scalar})
and~(\ref{eq:gauge}) are presented and discussed below.

\begin{figure}
\begin{center}
\includegraphics[width=7.7cm]{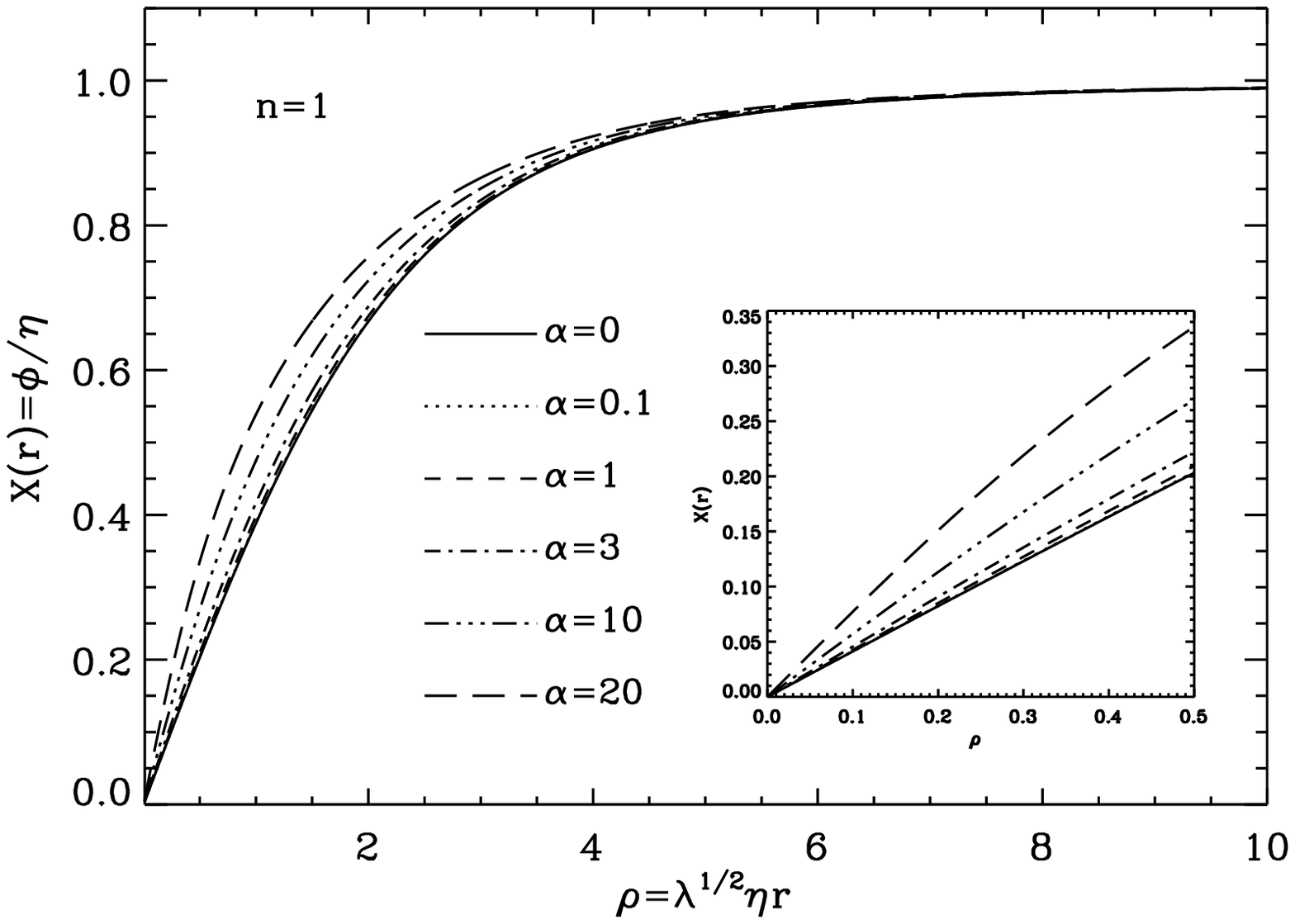}
\includegraphics[width=7.7cm]{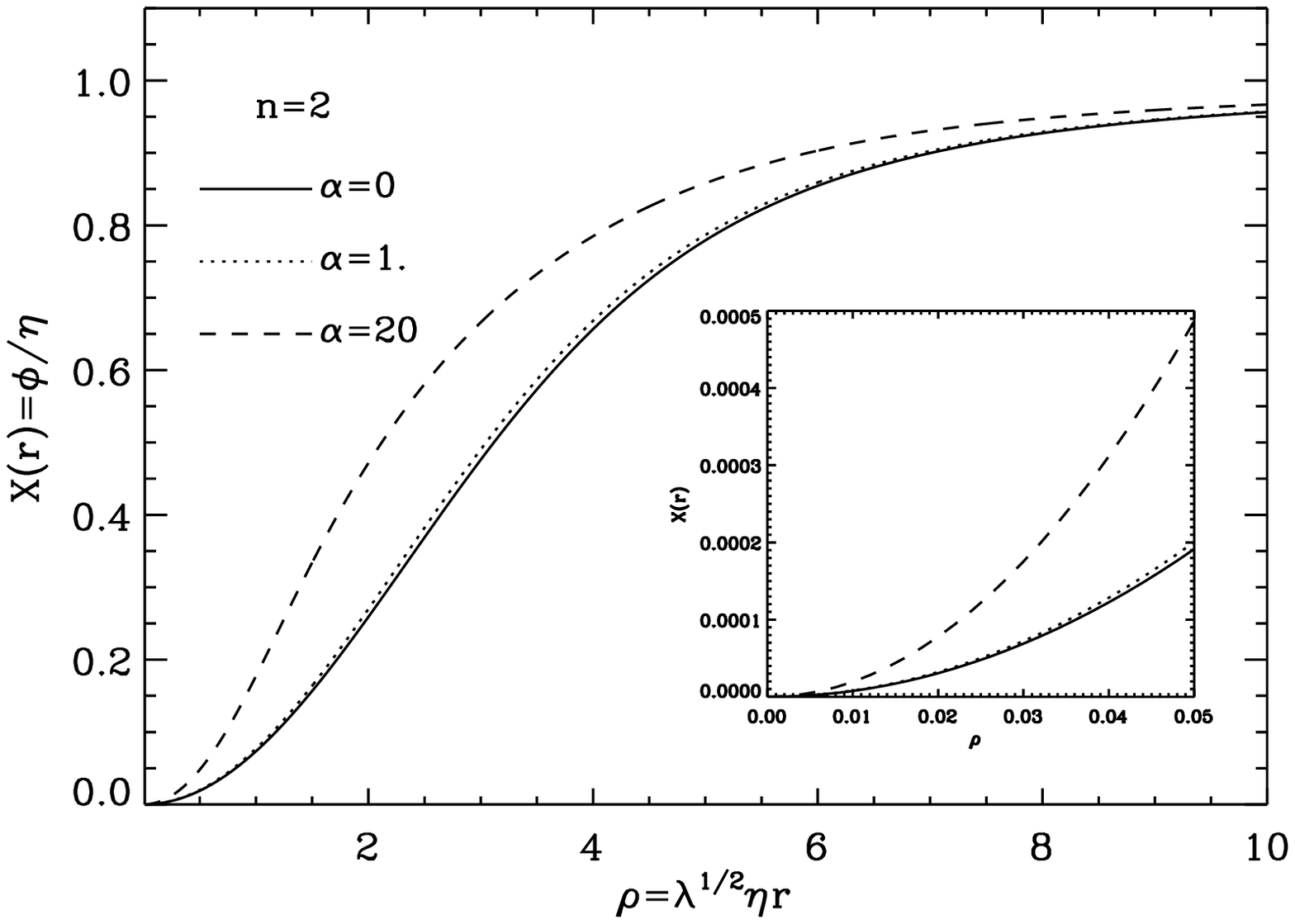}
\caption{Left panel: the profiles of global DBI strings for $n=1$ and
for various values of the parameter $\alpha$ defined in
Eq.~(\ref{eq:defalpha}). Right panel: same as left panel but with $n=2$}
\label{fig:globalprofile}
\end{center}
\end{figure}

Firstly, in Figs.~\ref{fig:globalprofile}, we consider global DBI
strings (that is to say without the gauge field) for, respectively,
winding numbers $n=1$ (left panel) and $n=2$ (right panel) and different
$\alpha $'s. This figure confirms the qualitative statements made in the
previous subsection. We notice that, even for ``non-perturbative''
values of $\alpha $, \ie $\alpha >1$, the difference between the
standard and the DBI profiles remains quite small. Moreover, as
announced, the maximum difference lies at a (dimensionless) radius
$\rho$ of order one, namely half way from the origin and the region
where $X\to 1$. Another remark is that the DBI profiles are always above
the standard profiles. This is of course expected since the DBI regime
means larger derivatives which, in the present context, implies the
above mentioned property. Finally, one can check that the asymptotic
behaviours discussed in the previous subsection are clearly observed in
Figs.~\ref{fig:globalprofile}. Indeed, for $n=1$, we notice that
$X(\rho)\sim A_{_{\rm DBI}}\rho$ where $A_{_{\rm DBI}}$ is clearly a
function of $\alpha$ (see in particular the zoom in the left panel). The
same remark applies for $n=2$, where $X(\rho)\sim \rho ^2$.

\begin{figure}
\begin{center}
\includegraphics[width=7.7cm]{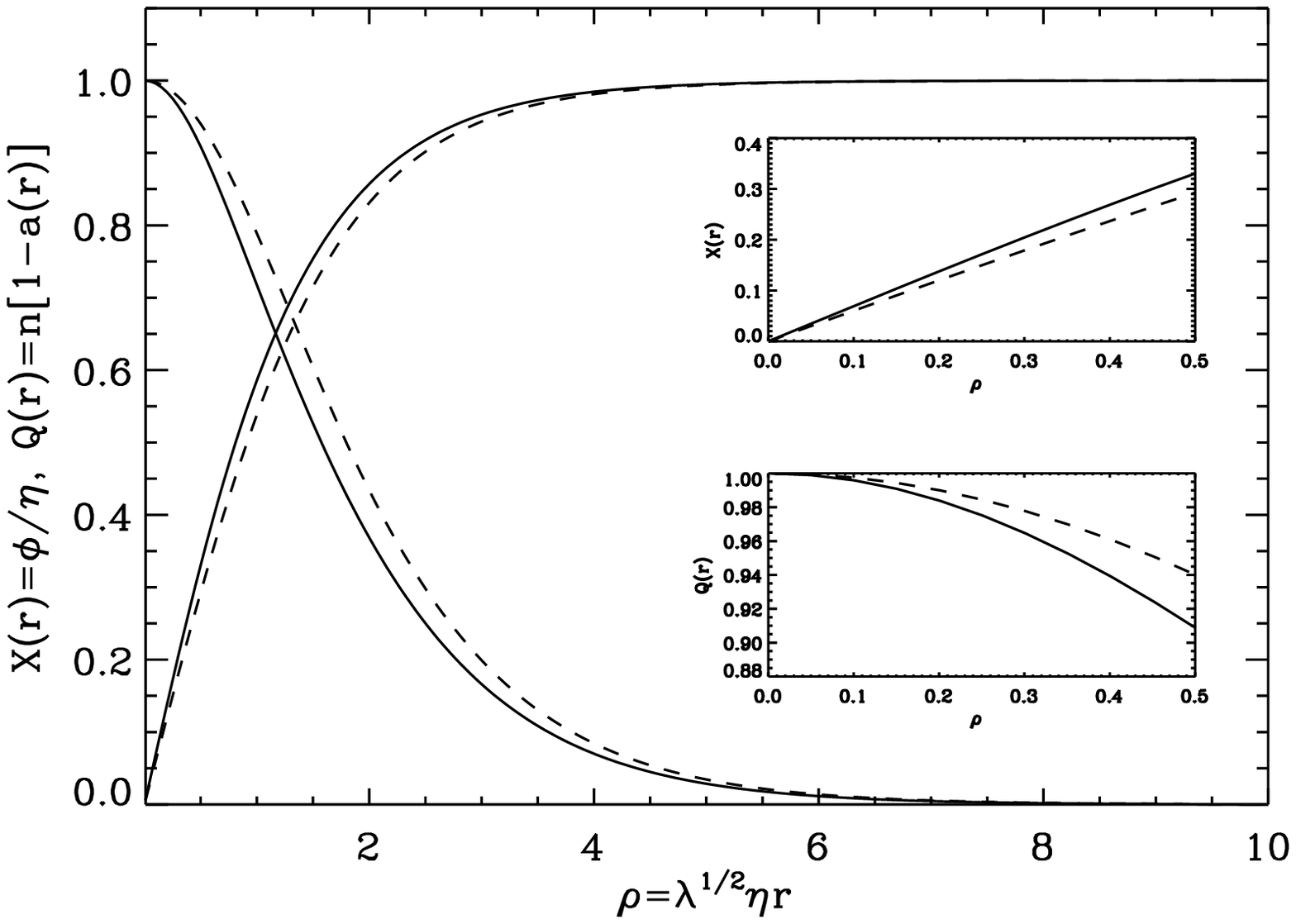}
\includegraphics[width=7.7cm]{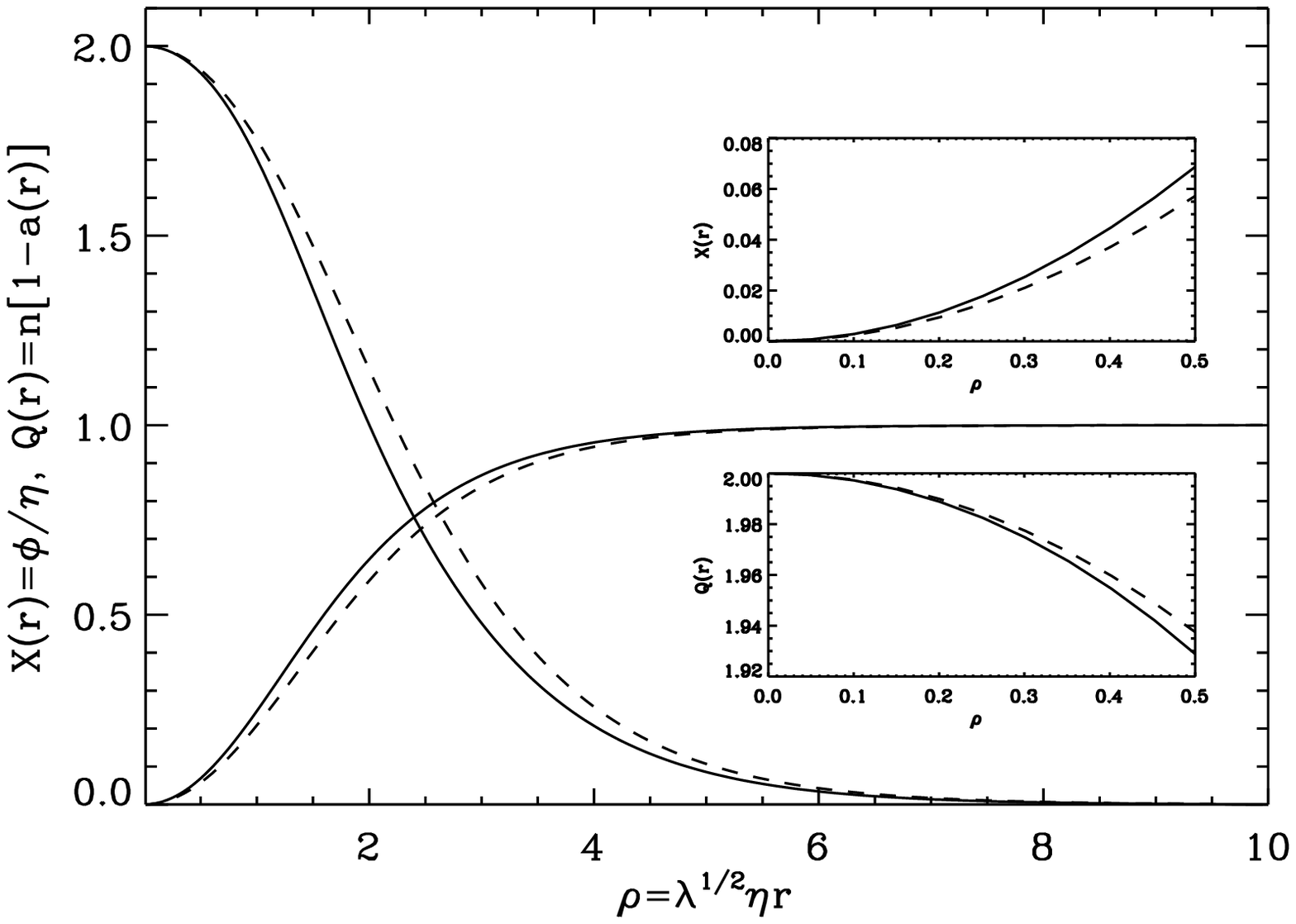} \caption{Left panel:
the solid lines represent the profiles of the scalar and gauge fields of
a DBI strings with $n=1$ and $\alpha=1$ while the dashed lines are the
profiles of the scalar and gauge fields of a standard Abelian-Higgs
string (\ie $\alpha=0$). Right panel: same as left panel but with
$n=2$. Notice that, on the $y$-axis, we have used the notation $a\equiv
\hat{q}\hat{A}_{\theta}/n$.}  \label{fig:dbiprofileal=1}
\end{center}
\end{figure}

Secondly, in Figs.~\ref{fig:dbiprofileal=1}, we display the profiles for
DBI local strings, \ie for the scalar field and the gauge field, in the
case where $\alpha =1$, $n=1$ (left panel) and $n=2$ (right panel). The
same remarks as before apply. In particular, the scalar field profile
always lies above its Abelian-Higgs counter part and, on the contrary,
the DBI gauge field profile always lies inside the standard profile. As
already discussed, this is because, in the DBI regime, the gradients
are, by definition, larger than in the standard case. This means that a
DBI string has a core smaller than an Abelian-Higgs string. As before,
one can also check that the asymptotic behaviours are those discussed in
the previous subsection. This is true in particular for the gauge field
for which we always see that $Q\sim n-\rho ^2$ at the origin.

\begin{figure}
\begin{center}
\includegraphics[width=7.7cm]{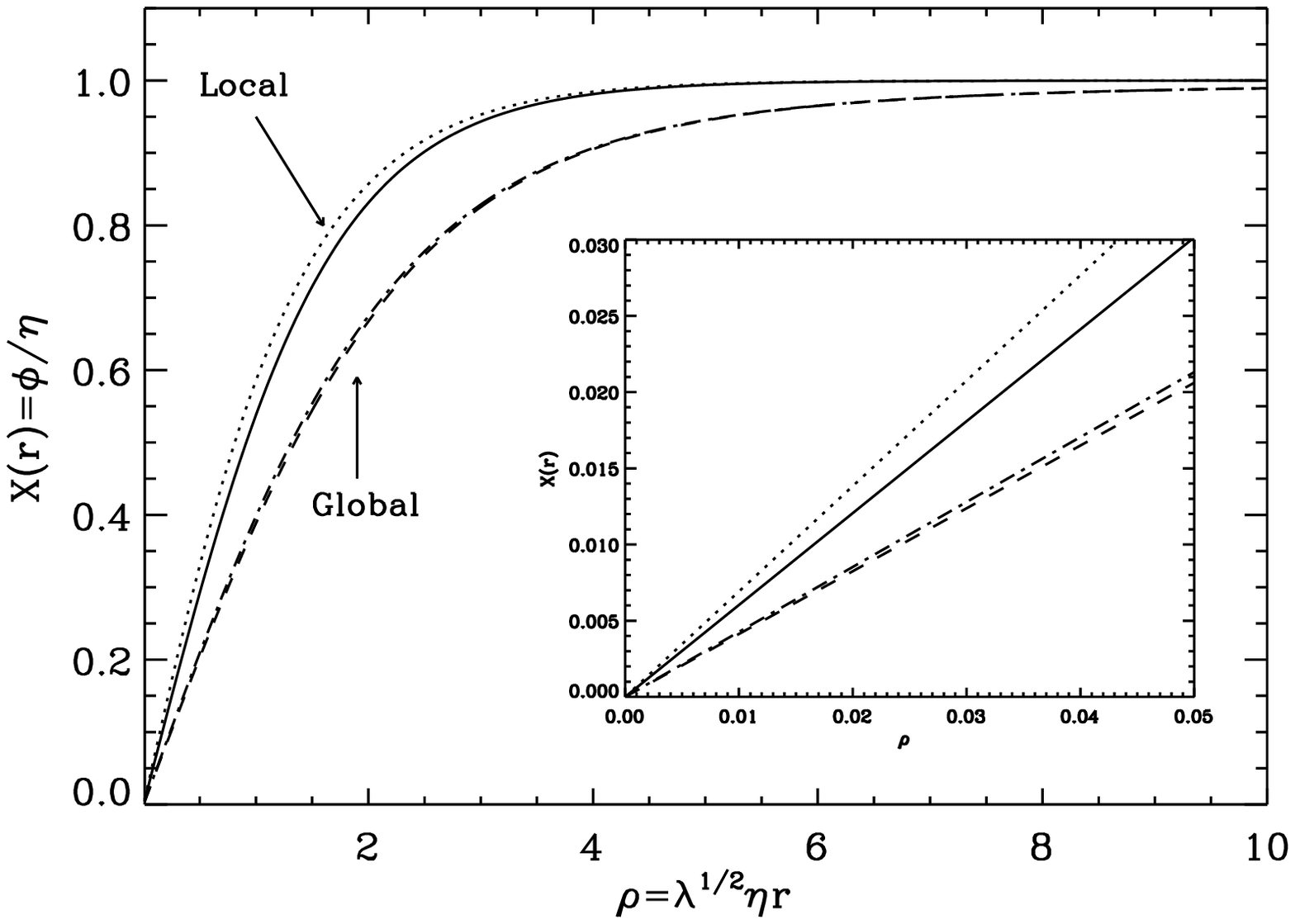}
\includegraphics[width=7.7cm]{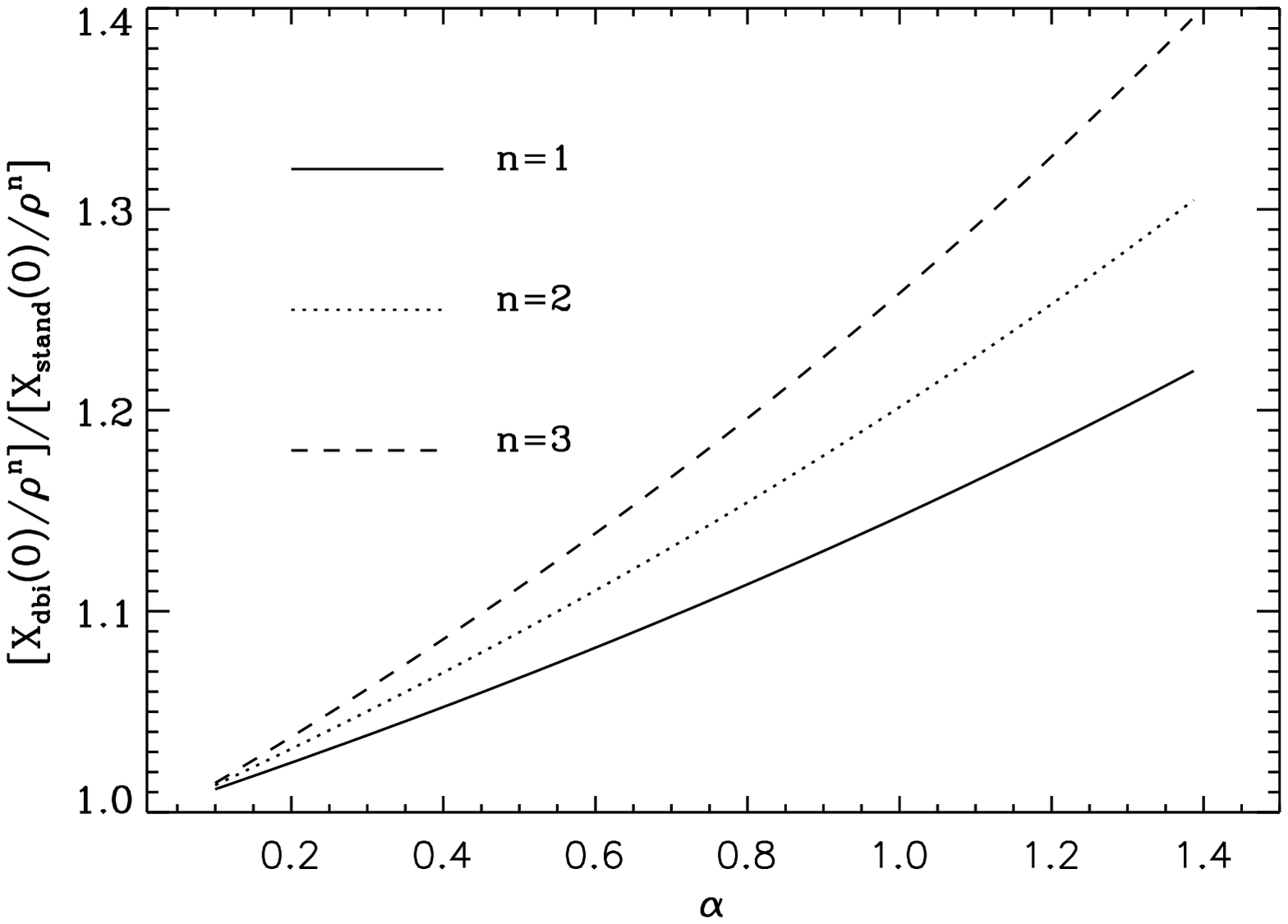} \caption{Left panel: comparison
of the cosmic string profiles for global and local DBI strings with
$n=1$ and $\alpha=1$. The solid line represents the scalar field profile
for a local Abelian-Higgs string while the dotted line is the
corresponding DBI profile. On the other hand, the dashed line represents
the scalar field profile for an Abelian-Higgs global string whereas the
dotted-dashed line is the corresponding DBI profile still in the global
case. Right panel: ratio $A_{_{\rm DBI}}/A_{_{\rm standard}}$ (see the
text) as a function of the parameter $\alpha$ for different values of
the winding number $n$.}  \label{fig:moreprofiles}
\end{center}
\end{figure}

Thirdly, additional information on the profiles can be gained from
Figs.~\ref{fig:moreprofiles}. In the left panel, we have compared the
local and global profiles. One can notice that the global profile is
less concentrated than the local one. Another remark is that the
difference between the Abelian-Higgs and DBI profiles is more important
in the local case than in the global one. In the right panel, we have
compared the slopes at the origin. In the standard case, one has
$X(\rho)\sim A_{_{\rm standard}}\rho ^n$ and it has been argued before
that in the DBI case, one also has $X(\rho)\sim A_{_{\rm DBI}}\rho
^n$. We have represented the ratio $A_{_{\rm DBI}}/A_{_{\rm standard}}$
for various values of $\alpha $ and $n$. One notices that the larger
$\alpha $, the steeper the DBI slope, which seems natural since the
value of the parameter $\alpha $ controls how important the DBI effects
are. We also remark that the same trend is observed when one increases
$n$ rather than $\alpha$. In conclusion, from these two figures, one
confirms that the deeper one penetrates into the DBI regime, the
narrower the core of a cosmic string is. The effect is larger in the
local case than in the global one and for large winding numbers than for
small ones.

\par

\begin{figure}
\begin{center}
\includegraphics[width=7.7cm]{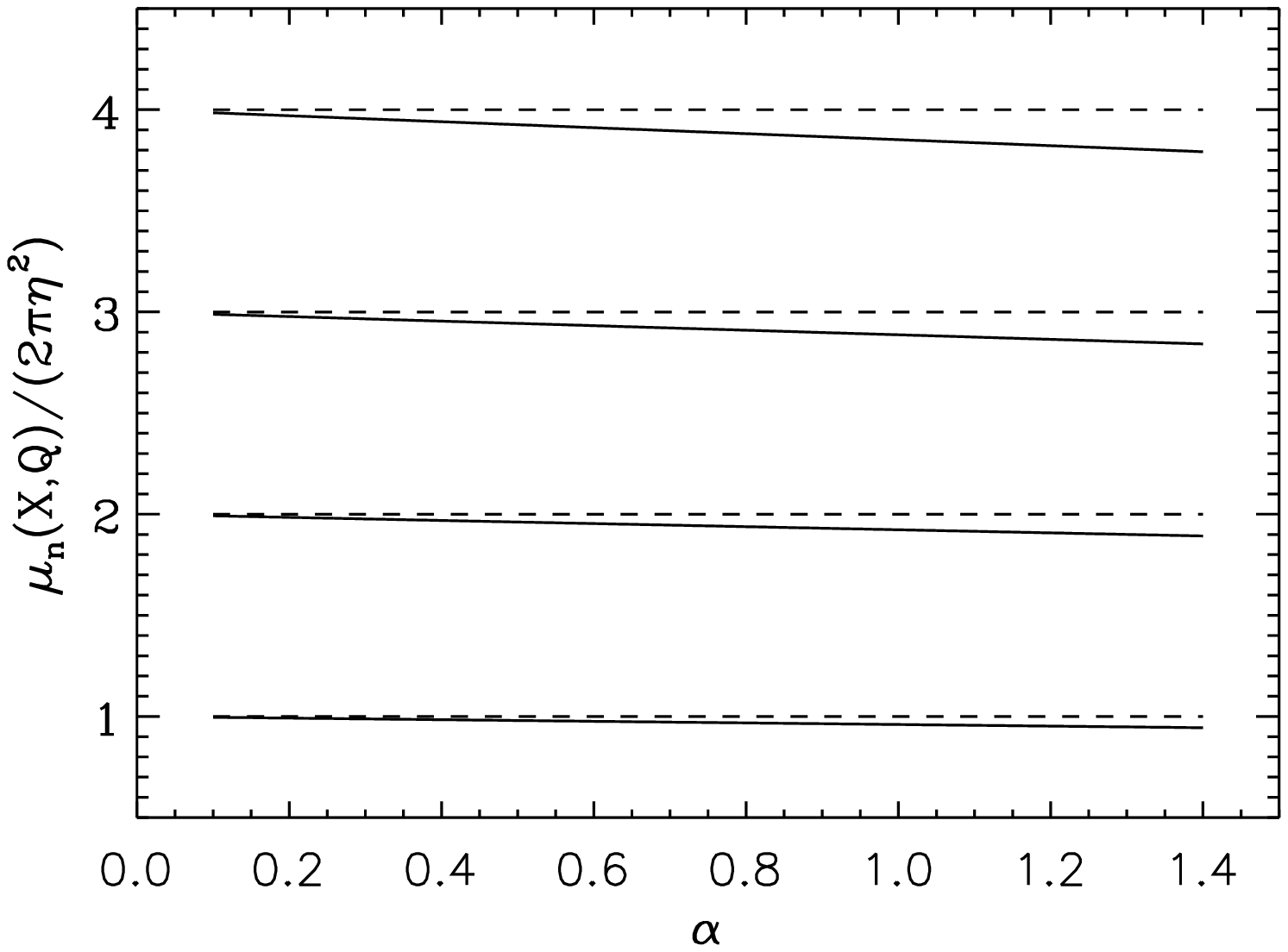}
\includegraphics[width=7.7cm]{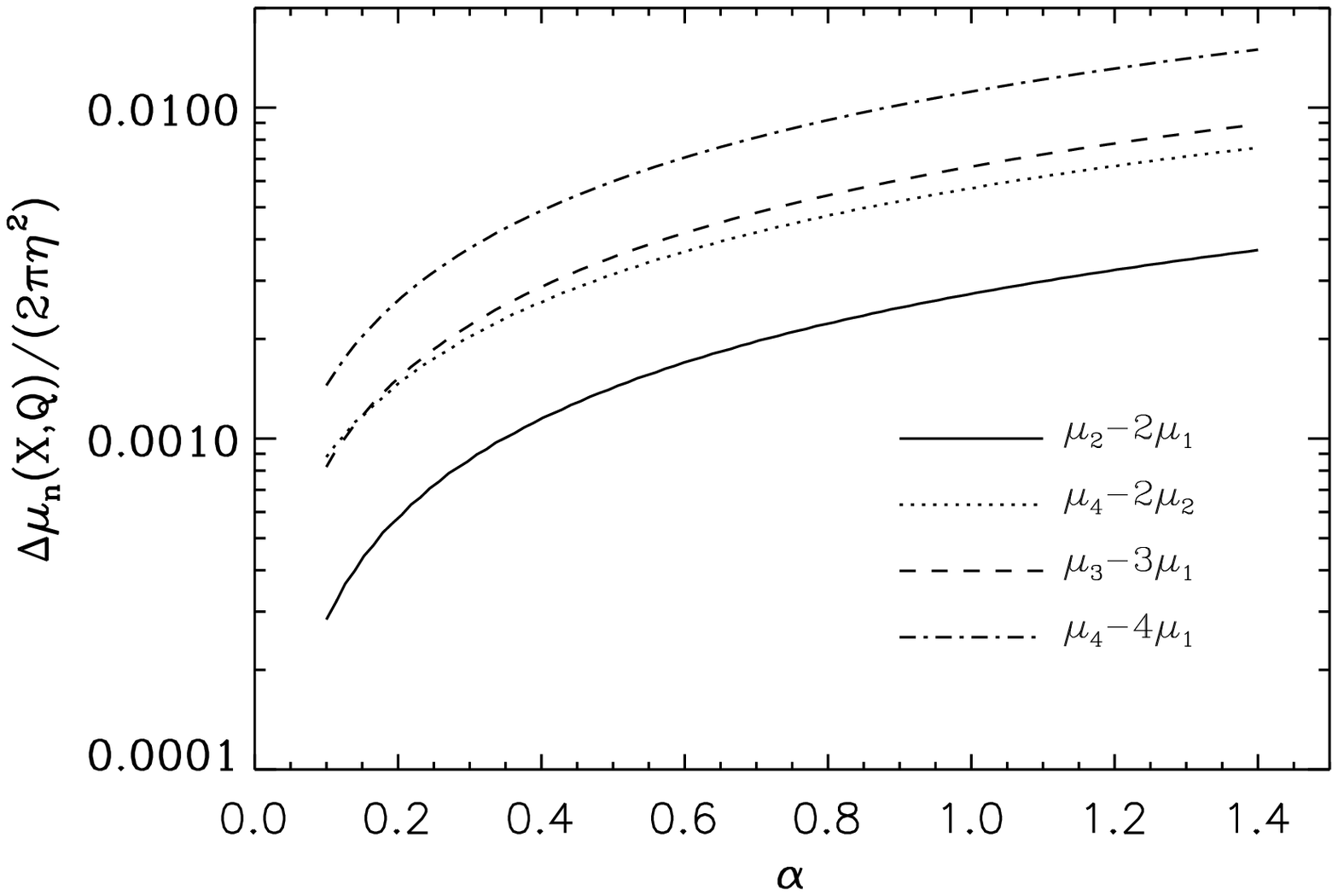} \caption{Left panel: the
solid lines represent the DBI string tension as a function of $\alpha$
for different values of the winding number $n$ (from $n=1$ to $n=4$
going from the bottom to the top of the plot). The dashed lines
corresponds to the Abelian-Higgs tension, namely $\mu _n=2\pi \eta ^2
n$, and are thus horizontal lines located at the $y$-coordinate
$n$. Right panel: the DBI string binding energy $\mu_{2n}-2\mu_n$ for
various $n$ as a function of the parameter $\alpha$.}
\label{fig:actiondbi}
\end{center}
\end{figure}

Fourthly, given the numerical solutions presented above it is
straightforward to calculate their tension which, from the action given
by Eq.~(\ref{eq:genten}), takes the form
\begin{equation}
\mu _n\left(X,Q\right)= 2\pi \eta^2 f_n(\alpha)\, ,
\end{equation}
where $f_n(0)=n$ in the Abelian-Higgs case. In Fig.~\ref{fig:actiondbi}
(left panel), we plot the universal functions $f_n(\alpha)$ for the DBI
local strings. We notice that the DBI action is, for any $n$ and/or
$\alpha $, smaller than the corresponding standard action. Moreover, at
a fixed value of $\alpha$, the (absolute) difference between the DBI and
Abelian-Higgs actions increases with the winding number. The fact that
the DBI action is smaller than the standard one is likely to have
important physical consequences, in particular with regards to the
formation of DBI cosmic strings. Indeed, if their energy is smaller than
in the standard case, one can legitimely expect their formation to be
favoured as compared to the Abelian-Higgs case.

In the right panel in Fig.~\ref{fig:actiondbi}, we have represented the
DBI string binding energy $\mu_{2n}-2\mu _n$ as a function of the
parameter $\alpha$ for different values of the winding number $n$. We
observe that this quantity is always positive but small in comparison to
one. Moreover, as expected since one has $\left(\mu_{2n}-2\mu
_n\right)\left(\alpha =0\right)=0$, it increases with $\alpha$. We
conclude that when $\alpha \neq 0$, the DBI cosmic string is no longer a
BPS object. The fact that $\mu_{2n}>2\mu _n$ means that, when they meet,
two DBI strings will not constitute a new single string with winding
number $2n$ since this appears to be disfavoured from the energy point
of view. This has important consequences for cosmology since the above
discussion implies that the behaviour of a network of DBI cosmic strings
will be similar to the behaviour of a network of Abelian-Higgs
strings. This means that the cosmological constraints derived, for
instance in Refs.~\cite{Pogosian:2003mz}, also apply to the present
case.

\par

\begin{figure}
\begin{center}
\includegraphics[width=7.7cm]{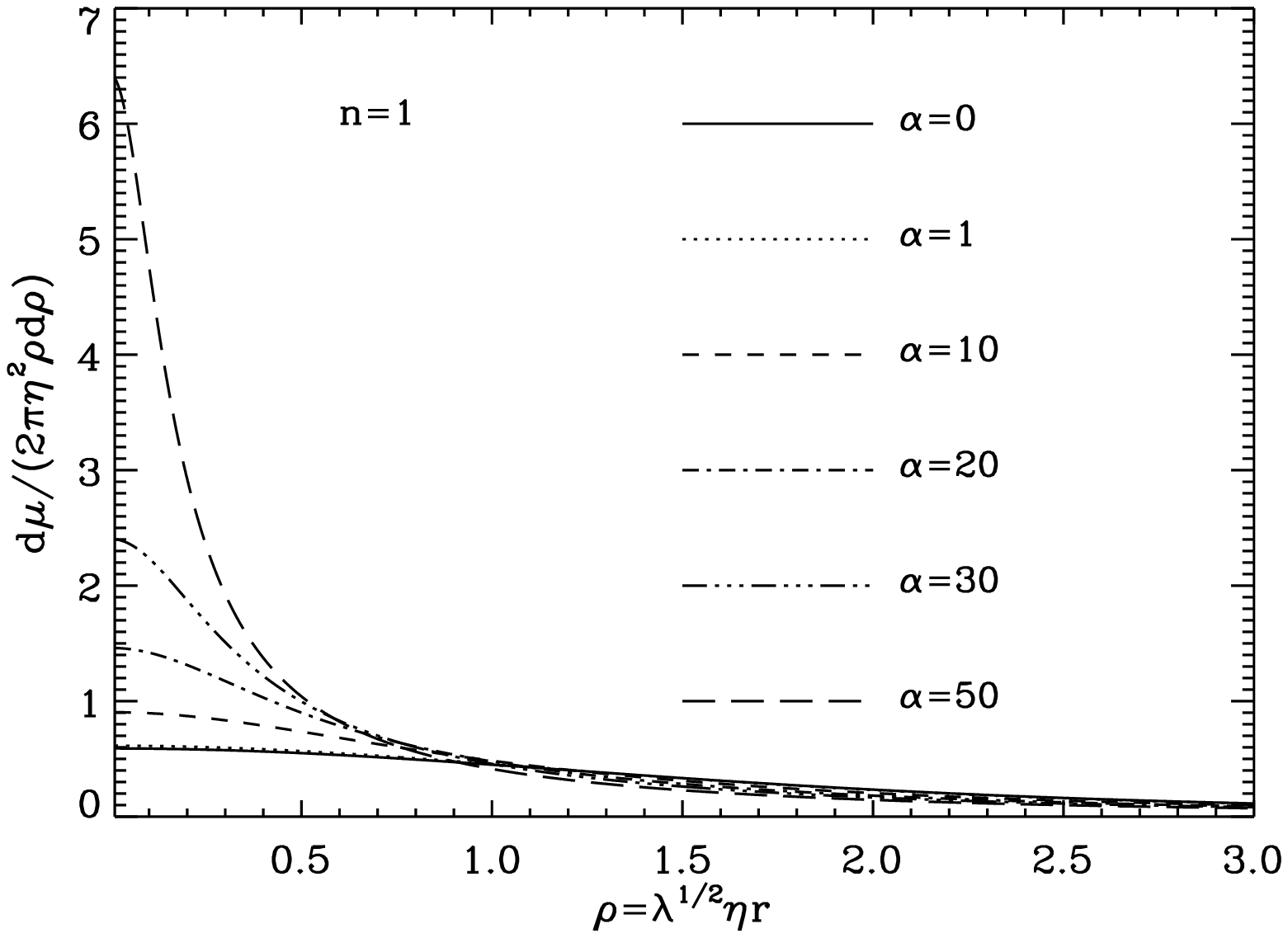}
\includegraphics[width=7.7cm]{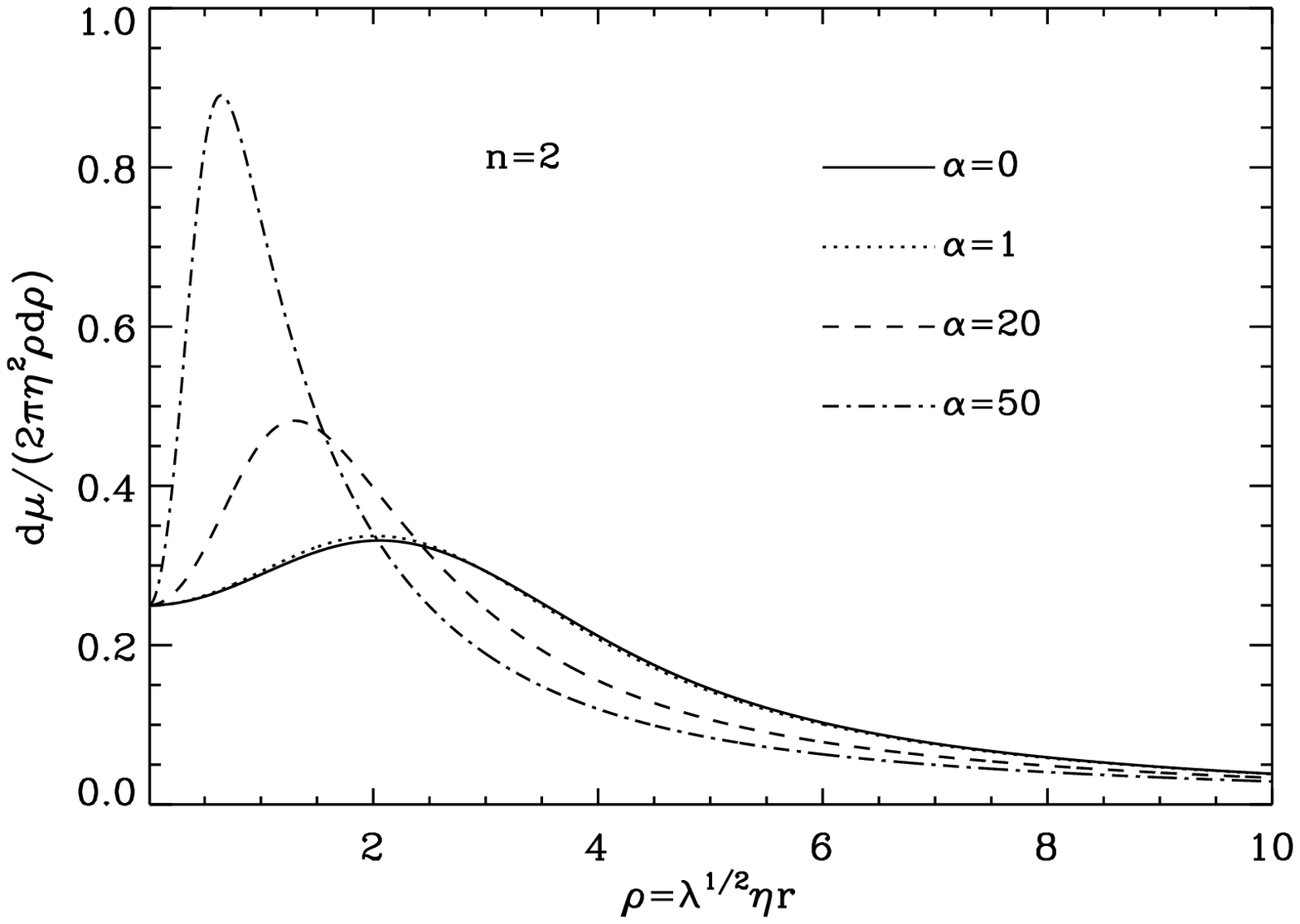} \caption{Left panel: the
energy density for DBI strings as a function of $\rho$ for $n=1$ for
different values of the parameter $\alpha$. Right panel: same as right
panel but for $n=2$.}
\label{fig:energydensity}
\end{center}
\end{figure}

Finally, in Fig.~\ref{fig:energydensity}, we represent the energy
density as a function of the dimensionless radial coordinate $\rho$ for
$n=1$ (left panel) and $n=2$ (right panel) for different values of the
parameter $\alpha $. We notice that the DBI energy densities are usually
more peaked than the Abelian-Higgs ones. Moreover, the larger $\alpha $,
the more peaked the distributions. The case $n=2$ is particularly
interesting. One observes that, as $\alpha $ increases, the peaks of the
distribution are displaced towards the left, \ie towards smaller values
of $\rho$. This is probably due to the fact that, as discussed at the
beginning of this subsection, the difference between the DBI and
Abelian-Higgs profiles is maximum for intermediate values of $\rho$.

\section{Conclusions}
\label{sec:conclusions}

We have considered a natural DBI generalisation of the Abelian-Higgs
model whereby the kinetic terms of the Higgs fields do not lead to a
linear differential operator in the equations of motion. The particular
form of the action is motivated by a specific extra-dimensional model
where the Higgs field becomes a complex direction normal to a
D3-brane. Although this model leads to nice cosmic string properties, it
is not directly related to a string theory model. As such, the closest
model of string theory which could have lead to such a DBI action is the
D3/D7 system where BPS cosmic strings are formed at the end of an
hybrid-like inflation phase. Unfortunately, the charged fields
associated to the open string joining the D3- and D7-branes have no
obvious geometric meaning and therefore do not lead to our DBI
action. It would certainly be very interesting to see if our
construction can be embedded within string theory.

\par

As a four-dimensional model of non-canonical type, the DBI model of
cosmic strings does not suffer from any pathology such as divergences or
non-single-valuedness of the field profiles (typical of other non-linear
actions which have been proposed in the literature). Indeed we find that
DBI strings can be continuously deformed to their Abelian-Higgs
analogue. In fact, the main difference from the Abelian-Higgs case
appears in the BPS case where the DBI strings show a small departure
from the BPS property. In particular, we find that the string tension is
reduced, a property which may have some phenomenological significance in
order to relax the bound on the string tension coming from Cosmic
Microwave Background (CMB) data. Moreover we find that the DBI strings
have a positive binding energy implying that the string coalescence is
energetically disfavoured, leading to the likely formation of networks
with singly-wound strings and statistical properties akin to the usual
Abelian-Higgs ones.

\par

In the present article, we have not tackled some important aspects of
DBI string dynamics such as string scattering (for which we expect that
the higher order terms discussed in the Appendix may play an important
r\^ole), gravitational back reaction, and the coupling to fermions and
their zero modes. This is currently under investigation.

\par

In summary, we have introduced DBI cosmic strings as non-singular
solutions derived from a non-linear Lagrangian. We have studied the
solutions numerically and found that they differ significantly from
their Abelian-Higgs analogues.  However, the network properties of these
strings is almost certainly similar to those of type II Abelian-Higgs
cosmic strings.

\acknowledgments

We would like to thank Patrick Peter and Christophe Ringeval for useful
discussions, particularly on the numerical aspect of this work. We also
thank David Langlois and S\'ebastien Renaux-Petel, as well as Renata
Kallosh for enlightening discussions on the stringy aspect of this work.

\section{Appendix}
\label{sec:appendix}

In this appendix, we study in more detail the general form of the
action~(\ref{eq:DBIaction}) considered in this
article. Eq.~(\ref{eq:DBIaction}) reads
\begin{eqnarray}
S &=& -T \int {\rm d}^4x \left\{\sqrt{-\det \left[ g_{\mu \nu} +
({\cal D}_{ ( \mu } \Phi) ({\cal D}_{\nu)} \Phi)^{\dagger} + \el^2 {\cal F}_{\mu\nu}
\right]}\nonumber \right. \\
& & \left. \vphantom{\sqrt{-\det \left[ g_{\mu \nu} +
({\cal D}_{ ( \mu } \Phi) ({\cal D}_{\nu)} \Phi)^{\dagger} +
(D_{\nu} \Phi) (D_{\mu} \Phi)^\dagger + \el^2 {\cal F}_{\mu\nu}
\right]}}
-\sqrt{-g}+\sqrt{-g}\frac{V\left(\sqrt{T}\left|\Phi\right|\right)}{T}
\right\}\, ,
\nonumber \\
&\equiv &-T \int {\rm d}^4 x \sqrt{-g}
\left[ \left(\sqrt{{\cal D}} -1 \right) +
\frac{V(\sqrt{T}|\Phi|)}{T} \right] \, ,
\label{bliss}
\end{eqnarray}
where ${\cal D}$ is defined by
\begin{equation}
{\cal D} \equiv \det \left[
\delta^{\mu}_{\;\; \nu} + ({\cal D}^{\mu} \Phi) ({\cal D}_{\nu}
\Phi)^{\dagger} + ({\cal D}^{\mu} \Phi)^\dagger ({\cal D}_{\nu}
\Phi) + \ells^2 {\cal F}^{\mu}{}_{\nu} \right]\, ,
\label{Ddef}
\end{equation}
 $T$ has dimensions of (energy)$^4$ and $\ells$ is a length scale. As
before, ${\cal D}_\mu= \partial_\mu -i\hat{q}{\cal A}_\mu$.

\par

Our goal is to compute and simplify Eq.~(\ref{Ddef}) for ${\cal D}$. As
it is clear from its definition, this will allow us to derive a more
compact formula for our action in the general case. In
Eq.~(\ref{eq:genten}) we have evaluated the action~(\ref{bliss}) for a
cylindrically symmetric static string profile.  In this case it takes a
simple form. However, when there is time dependence and less symmetry
--- as occurs for example in string scattering --- it is important to
know the general form of the action.

\par

First define the following quantities
\begin{equation}
N^{\nu} \equiv {\cal D}^{\mu} \Phi \, ,
\label{Ndef}
\quad
S^{\mu}{}_{\nu}\equiv N^{\mu}\bar{N}_\nu +\bar{N}^\mu
N_{\nu}\, ,
\label{Sdef}
\quad
R^{\mu}{}_{\nu}\equiv S^{\mu}{}_{\nu}+ {\cal F}^{\mu}{}_{\nu}\, ,
\label{Rdef}
\end{equation}
where are bar denotes complex conjugation and we set $\el=1$ in this
appendix. Note that by definition $S^{\mu\nu}$ is a symmetric matrix and
${\cal F}^{\mu\nu}$ is antisymmetric, while $S^{\mu}{}_{\nu}$ and ${\cal
F}^{\mu}{}_{\nu}$ are in general neither symmetric nor antisymmetric.
Denote by $S$ the matrix with components $S^{\mu}{}_{\nu}$, while ${\cal
F}$ is the matrix with components ${\cal F}^{\mu}{}_{\nu}$. For integer
$n$ and $p$
\begin{equation}
\tr \left(S^p{\cal F}^{2n+1}\right)=0\, .
\label{superU}
\end{equation}
On the other hand, we also have
\begin{eqnarray}
{\cal D} &=&
\det\left(\delta^{\mu}{}_{\nu}+R^{\mu}{}_{\nu}\right)\,
\\&=&
-\frac{1}{4!}\varepsilon_{\a_1\a_2\a_3\a_4}
\varepsilon^{\b_1\b_2\b_3\b_4}
\left(\delta^{\a_1}{}_{\beta_1}+\R^{\a_1}{}_{\beta_1}\right)
\left(\delta^{\a_2}{}_{\beta_2}+\R^{\a_2}{}_{\beta_2}\right)
\nonumber \\ & & \times
\left(\delta^{\a_3}{}_{\beta_3}+\R^{\a_3}{}_{\beta_3}\right)
\left(\delta^{\a_4}{}_{\beta_4}+\R^{\a_4}{}_{\beta_4}\right)\,
\label{expand}
\end{eqnarray}
which, on using the identity
\begin{equation}
\varepsilon_{\a_1\a_2\a_3\a_4}\varepsilon^{\alpha_1\cdots
\alpha_j\b_{j+1}\cdots\b_4}=-\left(4-j\right)!j!
\delta^{[\beta_{j+1}}_{\a_{j+1}}{}^{\cdots}_{\cdots}
\delta^{\beta_4]}_{\a_4}\, ,
\end{equation}
gives
\begin{equation}
{\cal D}= 1
+R^\alpha{}_{\alpha}+R^{[\alpha}{}_{\alpha}R^{\beta]}{}_{\beta}
+R^{[\alpha}{}_{\alpha}R^{\beta}{}_{\beta}R^{\gamma]}{}_{\gamma}
+R^{[\alpha}{}_{\alpha}R^{\beta}{}_{\beta}R^{\gamma}{}_{\gamma}
R^{\delta]}{}_{\delta}\, .
\end{equation}

We now evaluate each term in the above equation. For the
first (linear in $R$), it follows from Eqs.~(\ref{Sdef})
and~(\ref{superU}) that
\begin{equation}
R^\alpha{}_{\alpha}= S^\alpha{}_{\alpha}=2\bar{N}_\alpha N^\alpha =
2\left({\cal D}^{\mu}\Phi \right)\left({\cal D}_{\mu}
\Phi\right)^{\dagger}\, .
\end{equation}
The quadratic term is given by
\begin{eqnarray}
R^{[\alpha}{}_{\alpha}R^{\beta]}{}_{\beta}
&=& S^{[\alpha}{}_{\alpha}S^{\beta]}{}_{\beta}+
2S^{[\alpha}{}_{\alpha}{\cal F}^{\beta]}{}_{\beta}
+{\cal F}^{[\alpha}{}_{\alpha}{\cal F}^{\beta]}{}_{\beta}\, ,
\\&=&
\frac{1}{2}\left[\tr ^2 \left(S\right)- {\rm tr}\left(S^2\right)
\right]-\frac{1}{2}\tr\left({\cal F}^2\right)\, ,
\label{version1}
\\
&=&\left(\bar{N}_\alpha N^\alpha \right)^2
- \left({N}_{\alpha}N^{\alpha}\right)
\left(\bar{N}_\beta \bar{N}^\beta\right)
-\frac{1}{2}\tr\left({\cal F}^2\right)\, ,
\label{version2}
\end{eqnarray}
where to get from Eq.~(\ref{version1}) to Eq.~(\ref{version2}) we have
used Eq.~(\ref{Sdef}). Notice that these terms are compatible with the
$U(1)$ invariance of the action. The next step is to calculate the cubic
term. It is given by
\begin{eqnarray}
R^{[\alpha}{}_{\alpha}
R^{\beta}{}_{\beta}R^{\gamma]}{}_{\gamma}
&=&
S^{[\alpha}{}_{\alpha}S^{\beta}{}_{\beta}S^{\gamma]}{}_{\gamma}
+3 S^{[\alpha}{}_{\alpha}S^{\beta}{}_{\beta}{\cal
F}^{\gamma]}{}_{\gamma}
+3S^{[\alpha}{}_{\alpha}{\cal F}^{\beta}{}_{\beta}{\cal
F}^{\gamma]}{}_{\gamma}
\nonumber \\ & &
+ {\cal F}^{[\alpha}{}_{\alpha}{\cal
F}^{\beta}{}_{\beta}{\cal F}^{\gamma]}{}_{\gamma}\, .
\end{eqnarray}
The term in $S^3$ vanishes for the single complex scalar field studied
here since, on using Eq.~(\ref{Sdef}), it contains the contraction of an
antisymmetric tensor with a symmetric one. Similarly
$S^{[\alpha}{}_{\alpha}S^{\beta}{}_{\beta}{\cal F}^{\gamma]}{}_{\gamma}
=0={\cal F}^{[\alpha}{}_{\alpha}{\cal F}^{\beta}{}_{\beta} {\cal
F}^{\gamma]}{}_{\gamma}$ on using Eq.~(\ref{superU}).  Therefore, the
cubic term takes the form
\begin{equation}
R^{[\alpha}{}_{\alpha}R^{\beta}{}_{\beta}
R^{\gamma]}{}_{\gamma}
=3S^{[\alpha}{}_{\alpha}{\cal F}^{\beta}{}_{\beta}
{\cal F}^{\gamma]}{}_{\gamma}=\frac{1}{2}\left[ -\tr \left(S\right)
\tr\left({\cal F}^2\right)+2\tr\left(S{\cal F}^2\right)\right]\, .
\end{equation}

Finally, the quartic term can be expressed as
\begin{eqnarray}
R^{[\alpha}{}_{\alpha}R^{\beta}{}_{\beta}R^{\gamma}{}_{\gamma}
R^{\delta]}{}_{\delta}
&=&
S^{[\alpha}{}_{\alpha}S^{\beta}{}_{\beta}S^{\gamma}{}_{\gamma}
S^{\delta]}{}_{\delta}
+4S^{[\alpha}{}_{\alpha}S^{\beta}{}_{\beta}S^{\gamma}{}_{\gamma}
{\cal F}^{\delta]}{}_{\delta}
+4S^{[\alpha}{}_{\alpha}{\cal F}^{\beta}{}_{\beta}
{\cal F}^{\gamma}{}_{\gamma}{\cal F}^{\delta]}{}_{\delta}
\label{line1}
\nonumber  \\ & &
+6S^{[\alpha}{}_{\alpha}S^{\beta}{}_{\beta}
{\cal F}^{\gamma}{}_{\gamma} {\cal F}^{\delta]}{}_{\delta}
+{\cal F}^{[\alpha}{}_{\alpha}{\cal F}^{\beta}{}_{\beta}
{\cal F}^{\gamma}{}_{\gamma}{\cal F}^{\delta]}{}_{\delta}
\\
&=&
6S^{[\alpha}{}_{\alpha}S^{\beta}{}_{\beta}
{\cal F}^{\gamma}{}_{\gamma}{\cal F}^{\delta]}{}_{\delta}
+{\cal F}^{[\alpha}{}_{\alpha}{\cal F}^{\beta}{}_{\beta}
{\cal F}^{\gamma}{}_{\gamma}{\cal F}^{\delta]}{}_{\delta}\, ,
\label{line2}
\end{eqnarray}
since the terms on the first line in the above equations vanish, on
using the same arguments as above. Also
\begin{eqnarray}
S^{[\alpha}{}_{\alpha}S^{\beta}{}_{\beta}
{\cal F}^{\gamma}{}_{\gamma}{\cal F}^{\delta]}{}_{\delta}
&=&
\frac{1}{4!}\Biggl\{4\tr\left(S\right)
\tr \left(S{\cal F}^2\right)-4\tr\left({\cal F}^2 S^2\right)
-2\tr\left({\cal F}S{\cal F}S\right)
\nonumber
\\
& &+\left[\tr \left(S^2\right)
-\tr^2\left(S\right)\right]
\tr \left({\cal F}^2\right) \Biggr\}\, .
\\
{\cal F}^{[\alpha}{}_{\alpha}{\cal F}^{\beta}{}_{\beta}
{\cal F}^{\gamma}{}_{\gamma}{\cal F}^{\delta]}{}_{\delta}
&=& \frac{1}{4!}\left[-6\tr\left({\cal F}^4\right)
+3\tr^2 \left({\cal F}^2\right)\right]
\end{eqnarray}
Therefore, in the end, one obtains the following expression for ${\cal
D}$
\begin{eqnarray}
{\cal D} &=& 1+\tr\left(S\right)-\frac{1}{2}\tr\left({\cal
F}^2\right)
+\frac{1}{8}\left[\tr ^2\left({\cal F}^2\right)
-2\tr\left({\cal F}^4\right)\right]
\nonumber
\label{puke1}
\\
&& +\frac{1}{2} \left[\tr ^2\left(S\right)
-\tr\left(S^2\right)\right]
+ \frac{1}{2}\biggl[2\tr\left(S{\cal F}^2\right)
-\tr\left(S\right)\tr\left({\cal F}^2\right) \biggr]
\nonumber
\label{puke4}
\\
&&+ \frac{1}{4}\left[\tr\left(S^2\right)- \tr ^2\left(
S\right)\right] \tr\left({\cal F}^2\right)+
\tr\left(S\right)\tr\left(S{\cal F}^2\right) -\tr\left({\cal F}^2
S^2\right)
\label{ffinal}
\end{eqnarray}
The three terms of the first line in Eq.~(\ref{puke1}), when substituted
in Eq.~(\ref{bliss}) and on expanding the square-root, give the standard
Abelian-Higgs model.  The last two terms of Eq.~(\ref{puke1}) are the
standard terms of Born-Infeld electro-dynamics. Finally, as discussed in
the main text, the factor ${\cal D}$ and, hence, our action defined by
Eq.~(\ref{eq:DBIaction}), contains terms higher order in covariant
derivatives as well as mixing terms between ${\cal F}^2$ and the
covariant derivatives.

\section*{References}

\bibliography{referencedbics}

\providecommand{\href}[2]{#2}\begingroup\raggedright\begin{thebibliography}{10}

\bibitem{2008arXiv0803.0715G}
B.~{Gold}, C.~L. {Bennett}, R.~S. {Hill}, G.~{Hinshaw}, N.~{Odegard},
  L.~{Page}, D.~N. {Spergel}, J.~L. {Weiland}, J.~{Dunkley}, M.~{Halpern},
  N.~{Jarosik}, A.~{Kogut}, E.~{Komatsu}, D.~{Larson}, S.~S. {Meyer}, M.~R.
  {Nolta}, E.~{Wollack}, and E.~L. {Wright}, {\it {Five-Year Wilkinson
  Microwave Anisotropy Probe (WMAP) Observations: Galactic Foreground
  Emission}},  {\em ArXiv e-prints} {\bf 803} (Mar., 2008)
  [\href{http://xxx.lanl.gov/abs/0803.0715}{{\tt 0803.0715}}].

\bibitem{2008arXiv0803.0570H}
R.~S. {Hill}, J.~L. {Weiland}, N.~{Odegard}, E.~{Wollack}, G.~{Hinshaw},
  D.~{Larson}, C.~L. {Bennett}, M.~{Halpern}, L.~{Page}, J.~{Dunkley},
  B.~{Gold}, N.~{Jarosik}, A.~{Kogut}, M.~{Limon}, M.~R. {Nolta}, D.~N.
  {Spergel}, G.~S. {Tucker}, and E.~L. {Wright}, {\it {Five-Year Wilkinson
  Microwave Anisotropy Probe (WMAP)Observations: Beam Maps and Window
  Functions}},  {\em ArXiv e-prints} {\bf 803} (Mar., 2008)
  [\href{http://xxx.lanl.gov/abs/0803.0570}{{\tt 0803.0570}}].

\bibitem{2008arXiv0803.0732H}
G.~{Hinshaw}, J.~L. {Weiland}, R.~S. {Hill}, N.~{Odegard}, D.~{Larson}, C.~L.
  {Bennett}, J.~{Dunkley}, B.~{Gold}, M.~R. {Greason}, N.~{Jarosik},
  E.~{Komatsu}, M.~R. {Nolta}, L.~{Page}, D.~N. {Spergel}, E.~{Wollack},
  M.~{Halpern}, A.~{Kogut}, M.~{Limon}, S.~S. {Meyer}, G.~S. {Tucker}, and
  E.~L. {Wright}, {\it {Five-Year Wilkinson Microwave Anisotropy Probe (WMAP)
  Observations: Data Processing, Sky Maps, and Basic Results}},  {\em ArXiv
  e-prints} {\bf 803} (Mar., 2008)
  [\href{http://xxx.lanl.gov/abs/0803.0732}{{\tt 0803.0732}}].

\bibitem{2008arXiv0803.0593N}
M.~R. {Nolta}, J.~{Dunkley}, R.~S. {Hill}, G.~{Hinshaw}, E.~{Komatsu},
  D.~{Larson}, L.~{Page}, D.~N. {Spergel}, C.~L. {Bennett}, B.~{Gold},
  N.~{Jarosik}, N.~{Odegard}, J.~L. {Weiland}, E.~{Wollack}, M.~{Halpern},
  A.~{Kogut}, M.~{Limon}, S.~S. {Meyer}, G.~S. {Tucker}, and E.~L. {Wright},
  {\it {Five-Year Wilkinson Microwave Anisotropy Probe (WMAP) Observations:
  Angular Power Spectra}},  {\em ArXiv e-prints} {\bf 803} (Mar., 2008)
  [\href{http://xxx.lanl.gov/abs/0803.0593}{{\tt 0803.0593}}].

\bibitem{2008arXiv0803.0586D}
J.~{Dunkley}, E.~{Komatsu}, M.~R. {Nolta}, D.~N. {Spergel}, D.~{Larson},
  G.~{Hinshaw}, L.~{Page}, C.~L. {Bennett}, B.~{Gold}, N.~{Jarosik}, J.~L.
  {Weiland}, M.~{Halpern}, R.~S. {Hill}, A.~{Kogut}, M.~{Limon}, S.~S. {Meyer},
  G.~S. {Tucker}, E.~{Wollack}, and E.~L. {Wright}, {\it {Five-Year Wilkinson
  Microwave Anisotropy Probe (WMAP) Observations: Likelihoods and Parameters
  from the WMAP data}},  {\em ArXiv e-prints} {\bf 803} (Mar., 2008)
  [\href{http://xxx.lanl.gov/abs/0803.0586}{{\tt 0803.0586}}].

\bibitem{2008arXiv0803.0547K}
E.~{Komatsu}, J.~{Dunkley}, M.~R. {Nolta}, C.~L. {Bennett}, B.~{Gold},
  G.~{Hinshaw}, N.~{Jarosik}, D.~{Larson}, M.~{Limon}, L.~{Page}, D.~N.
  {Spergel}, M.~{Halpern}, R.~S. {Hill}, A.~{Kogut}, S.~S. {Meyer}, G.~S.
  {Tucker}, J.~L. {Weiland}, E.~{Wollack}, and E.~L. {Wright}, {\it {Five-Year
  Wilkinson Microwave Anisotropy Probe (WMAP) Observations: Cosmological
  Interpretation}},  {\em ArXiv e-prints} {\bf 803} (Mar., 2008)
  [\href{http://xxx.lanl.gov/abs/0803.0547}{{\tt 0803.0547}}].

\bibitem{inf25}
M.~{Lemoine}, J.~{Martin}, and P.~{Peter (Eds.)}, {\em {\sl Inflationary
  cosmology}, {\rm Lect. Notes Phys.} {\bf 738}}.
\newblock Springer, Berlin Heidelberg, 2008.

\bibitem{McAllister:2007bg}
L.~McAllister and E.~Silverstein, {\it {String Cosmology: A Review}},  {\em
  Gen. Rel. Grav.} {\bf 40} (2008) 565--605,
  [\href{http://xxx.lanl.gov/abs/0710.2951}{{\tt 0710.2951}}].

\bibitem{Burgess:2007pz}
C.~P. Burgess, {\it {Lectures on Cosmic Inflation and its Potential Stringy
  Realizations}},  {\em PoS} {\bf P2GC} (2006) 008,
  [\href{http://xxx.lanl.gov/abs/0708.2865}{{\tt 0708.2865}}].

\bibitem{Kallosh:2007ig}
R.~Kallosh, {\it {On Inflation in String Theory}},  {\em Lect. Notes Phys.}
  {\bf 738} (2008) 119--156,
  [\href{http://xxx.lanl.gov/abs/hep-th/0702059}{{\tt hep-th/0702059}}].

\bibitem{HenryTye:2006uv}
S.~H. Henry~Tye, {\it {Brane inflation: String theory viewed from the cosmos}},
   {\em Lect. Notes Phys.} {\bf 737} (2008) 949--974,
  [\href{http://xxx.lanl.gov/abs/hep-th/0610221}{{\tt hep-th/0610221}}].

\bibitem{Silverstein:2003hf}
E.~Silverstein and D.~Tong, {\it {Scalar speed limits and cosmology:
  Acceleration from D- cceleration}},  {\em Phys. Rev.} {\bf D70} (2004)
  103505, [\href{http://xxx.lanl.gov/abs/hep-th/0310221}{{\tt
  hep-th/0310221}}].

\bibitem{Alishahiha:2004eh}
M.~Alishahiha, E.~Silverstein, and D.~Tong, {\it {DBI in the sky}},  {\em Phys.
  Rev.} {\bf D70} (2004) 123505,
  [\href{http://xxx.lanl.gov/abs/hep-th/0404084}{{\tt hep-th/0404084}}].

\bibitem{Chen:2004gc}
X.~Chen, {\it {Multi-throat brane inflation}},  {\em Phys. Rev.} {\bf D71}
  (2005) 063506, [\href{http://xxx.lanl.gov/abs/hep-th/0408084}{{\tt
  hep-th/0408084}}].

\bibitem{Chen:2005ad}
X.~Chen, {\it {Inflation from warped space}},  {\em JHEP} {\bf 08} (2005) 045,
  [\href{http://xxx.lanl.gov/abs/hep-th/0501184}{{\tt hep-th/0501184}}].

\bibitem{Lorenz:2007ze}
L.~Lorenz, J.~Martin, and C.~Ringeval, {\it {Brane inflation and the WMAP data:
  a Bayesian analysis}},  {\em JCAP} {\bf 0804} (2008) 001,
  [\href{http://xxx.lanl.gov/abs/0709.3758}{{\tt 0709.3758}}].

\bibitem{Lorenz:2008je}
L.~Lorenz, J.~Martin, and C.~Ringeval, {\it {Constraints on Kinetically
  Modified Inflation from WMAP5}},
  \href{http://xxx.lanl.gov/abs/0807.2414}{{\tt 0807.2414}}.

\bibitem{Lorenz:2008et}
L.~Lorenz, J.~Martin, and C.~Ringeval, {\it {K-inflationary Power Spectra in
  the Uniform Approximation}},  \href{http://xxx.lanl.gov/abs/0807.3037}{{\tt
  0807.3037}}.

\bibitem{Langlois:2008qf}
D.~Langlois, S.~Renaux-Petel, D.~A. Steer, and T.~Tanaka, {\it {Primordial
  perturbations and non-Gaussianities in DBI and general multi-field
  inflation}},  \href{http://xxx.lanl.gov/abs/0806.0336}{{\tt 0806.0336}}.

\bibitem{Langlois:2008wt}
D.~Langlois, S.~Renaux-Petel, D.~A. Steer, and T.~Tanaka, {\it {Primordial
  fluctuations and non-Gaussianities in multi- field DBI inflation}},  {\em
  Phys. Rev. Lett.} {\bf 101} (2008) 061301,
  [\href{http://xxx.lanl.gov/abs/0804.3139}{{\tt 0804.3139}}].

\bibitem{Dasgupta:2002ew}
K.~Dasgupta, C.~Herdeiro, S.~Hirano, and R.~Kallosh, {\it {D3/D7 inflationary
  model and M-theory}},  {\em Phys. Rev.} {\bf D65} (2002) 126002,
  [\href{http://xxx.lanl.gov/abs/hep-th/0203019}{{\tt hep-th/0203019}}].

\bibitem{Dasgupta:2004dw}
K.~Dasgupta, J.~P. Hsu, R.~Kallosh, A.~Linde, and M.~Zagermann, {\it {D3/D7
  brane inflation and semilocal strings}},  {\em JHEP} {\bf 08} (2004) 030,
  [\href{http://xxx.lanl.gov/abs/hep-th/0405247}{{\tt hep-th/0405247}}].

\bibitem{Copeland:2003bj}
E.~J. Copeland, R.~C. Myers, and J.~Polchinski, {\it {Cosmic F- and
  D-strings}},  {\em JHEP} {\bf 06} (2004) 013,
  [\href{http://xxx.lanl.gov/abs/hep-th/0312067}{{\tt hep-th/0312067}}].

\bibitem{Pogosian:2003mz}
L.~Pogosian, S.~H.~H. Tye, I.~Wasserman, and M.~Wyman, {\it {Observational
  constraints on cosmic string production during brane inflation}},  {\em Phys.
  Rev.} {\bf D68} (2003) 023506,
  [\href{http://xxx.lanl.gov/abs/hep-th/0304188}{{\tt hep-th/0304188}}].

\bibitem{Dvali:2003zh}
G.~Dvali, R.~Kallosh, and A.~Van~Proeyen, {\it {D-term strings}},  {\em JHEP}
  {\bf 01} (2004) 035, [\href{http://xxx.lanl.gov/abs/hep-th/0312005}{{\tt
  hep-th/0312005}}].

\bibitem{Brax:2006zu}
P.~Brax, C.~van~de Bruck, A.~C. Davis, and S.~C. Davis, {\it {Cosmic D-strings
  and vortons in supergravity}},  {\em Phys. Lett.} {\bf B640} (2006) 7--12,
  [\href{http://xxx.lanl.gov/abs/hep-th/0606036}{{\tt hep-th/0606036}}].

\bibitem{Moreno:1998vy}
E.~Moreno, C.~Nunez, and F.~A. Schaposnik, {\it {Electrically charged vortex
  solution in Born-Infeld theory}},  {\em Phys. Rev.} {\bf D58} (1998) 025015,
  [\href{http://xxx.lanl.gov/abs/hep-th/9802175}{{\tt hep-th/9802175}}].

\bibitem{Yang:2000uj}
Y.~S. Yang, {\it {Classical solutions in the Born-Infeld theory}},  {\em Proc.
  Roy. Soc. Lond.} {\bf A456} (2000) 615--640.

\bibitem{Sarangi:2007mj}
S.~Sarangi, {\it {DBI Global Strings}},  {\em JHEP} {\bf 07} (2008) 018,
  [\href{http://xxx.lanl.gov/abs/0710.0421}{{\tt 0710.0421}}].

\bibitem{Brihaye:2001ag}
Y.~Brihaye and B.~Mercier, {\it {Classical solutions in the
  Einstein-Born-Infeld-Abelian- Higgs model}},  {\em Phys. Rev.} {\bf D64}
  (2001) 044001, [\href{http://xxx.lanl.gov/abs/hep-th/0102002}{{\tt
  hep-th/0102002}}].

\bibitem{Babichev:2006cy}
E.~Babichev, {\it {Global topological k-defects}},  {\em Phys. Rev.} {\bf D74}
  (2006) 085004, [\href{http://xxx.lanl.gov/abs/hep-th/0608071}{{\tt
  hep-th/0608071}}].

\bibitem{Babichev:2007tn}
E.~Babichev, {\it {Gauge k-vortices}},  {\em Phys. Rev.} {\bf D77} (2008)
  065021, [\href{http://xxx.lanl.gov/abs/0711.0376}{{\tt 0711.0376}}].

\bibitem{Hindmarsh:1994re}
M.~B. Hindmarsh and T.~W.~B. Kibble, {\it {Cosmic strings}},  {\em Rept. Prog.
  Phys.} {\bf 58} (1995) 477--562,
  [\href{http://xxx.lanl.gov/abs/hep-ph/9411342}{{\tt hep-ph/9411342}}].

\bibitem{Vilenkin:1994}
A.~Vilenkin and E.~P.~S. Shellard, {\em Cosmic Stringas and Other Topological
  Defects}.
\newblock Cambridge University Press, Cambridge, 1994.

\bibitem{Rubakov}
V.~Rubakov, {\em Classical theory of gauge fields}.
\newblock Cambridge University Press, Cambridge, 2002.

\bibitem{Bevis:2007gh}
N.~Bevis, M.~Hindmarsh, M.~Kunz, and J.~Urrestilla, {\it {Fitting CMB data with
  cosmic strings and inflation}},  {\em Phys. Rev. Lett.} {\bf 100} (2008)
  021301, [\href{http://xxx.lanl.gov/abs/astro-ph/0702223}{{\tt
  astro-ph/0702223}}].

\bibitem{Jacobs:1978ch}
L.~Jacobs and C.~Rebbi, {\it {Interaction Energy of Superconducting Vortices}},
   {\em Phys. Rev.} {\bf B19} (1979) 4486--4494.

\bibitem{Bettencourt:1994kc}
L.~M.~A. Bettencourt and T.~W.~B. Kibble, {\it {Nonintercommuting
  configurations in the collisions of type I U(1) cosmic strings}},  {\em Phys.
  Lett.} {\bf B332} (1994) 297--304,
  [\href{http://xxx.lanl.gov/abs/hep-ph/9405221}{{\tt hep-ph/9405221}}].

\bibitem{Copeland:2006eh}
E.~J. Copeland, T.~W.~B. Kibble, and D.~A. Steer, {\it {Collisions of strings
  with Y junctions}},  {\em Phys. Rev. Lett.} {\bf 97} (2006) 021602,
  [\href{http://xxx.lanl.gov/abs/hep-th/0601153}{{\tt hep-th/0601153}}].

\bibitem{Salmi:2007ah}
P.~Salmi {\em et~al.}, {\it {Kinematic Constraints on Formation of Bound States
  of Cosmic Strings - Field Theoretical Approach}},  {\em Phys. Rev.} {\bf D77}
  (2008) 041701, [\href{http://xxx.lanl.gov/abs/0712.1204}{{\tt 0712.1204}}].

\bibitem{Hsu:2003cy}
J.~P. Hsu, R.~Kallosh, and S.~Prokushkin, {\it {On brane inflation with volume
  stabilization}},  {\em JCAP} {\bf 0312} (2003) 009,
  [\href{http://xxx.lanl.gov/abs/hep-th/0311077}{{\tt hep-th/0311077}}].

\bibitem{Binetruy:2004hh}
P.~Binetruy, G.~Dvali, R.~Kallosh, and A.~Van~Proeyen, {\it {Fayet-Iliopoulos
  terms in supergravity and cosmology}},  {\em Class. Quant. Grav.} {\bf 21}
  (2004) 3137--3170, [\href{http://xxx.lanl.gov/abs/hep-th/0402046}{{\tt
  hep-th/0402046}}].

\bibitem{Burgess:2003ic}
C.~P. Burgess, R.~Kallosh, and F.~Quevedo, {\it {de Sitter string vacua from
  supersymmetric D-terms}},  {\em JHEP} {\bf 10} (2003) 056,
  [\href{http://xxx.lanl.gov/abs/hep-th/0309187}{{\tt hep-th/0309187}}].

\bibitem{sen}
A.~Sen, {\it {Dirac-Born-Infeld action on the tachyon kink and vortex}},  {\em
  Phys. Rev.} {\bf D68} (2003) 066008,
  [\href{http://xxx.lanl.gov/abs/hep-th/0303057}{{\tt hep-th/0303057}}].

\bibitem{Kim:2005tw}
Y.~Kim, B.~Kyae, and J.~Lee, {\it {Global and local D-vortices}},  {\em JHEP}
  {\bf 10} (2005) 002, [\href{http://xxx.lanl.gov/abs/hep-th/0508027}{{\tt
  hep-th/0508027}}].

\bibitem{Myers}
M.~R. Garousi and R.~C. Myers, {\it {World-volume interactions on D-branes}},
  {\em Nucl. Phys.} {\bf B542} (1999) 73--88,
  [\href{http://xxx.lanl.gov/abs/hep-th/9809100}{{\tt hep-th/9809100}}].

\bibitem{Born:1934gh}
M.~Born and L.~Infeld, {\it {Foundations of the new field theory}},  {\em Proc.
  Roy. Soc. Lond.} {\bf A144} (1934) 425--451.

\bibitem{AP84}
S.~L. {Adler} and T.~{Piran}, {\it Relaxation methods for gauge field
  equilibrium equations},  {\em Rev. Mod. Phys.} {\bf 56} (Jan., 1984) 1--40.

\bibitem{Peter:1992dw}
P.~Peter, {\it Superconducting cosmic string: Equation of state for space -
  like and time - like current in the neutral limit},  {\em Phys. Rev.} {\bf
  D45} (1992) 1091--1102.

\bibitem{Ringeval:2002qi}
C.~Ringeval, {\em Fermionic currents flowing along extended objects}.
\newblock Phd Thesis, University Paris 6, 2002.

\end{thebibliography}\endgroup

\end{document}